\documentclass[runningheads]{llncs}
\usepackage[T1]{fontenc}
\usepackage{graphicx}
\usepackage{paralist}
\usepackage{xcolor}
\usepackage{amsmath,amsfonts,amssymb}
\usepackage{subfiles}
\usepackage{bcprules}
\usepackage{bussproofs}
\usepackage{stackengine}
\usepackage{tabularx}
\usepackage{enumitem}
\usepackage{multirow}
\usepackage{algorithm}
\usepackage[noend]{algpseudocode}
\usepackage{fancyvrb}
\fvset{
  numbers=left,
  numbersep=2mm,
  commandchars=\\\{\}
}
\makeatletter
\RequirePackage[bookmarks,unicode,colorlinks=true]{hyperref}%
\usepackage{cleveref}
   \def\@citecolor{blue}%
   \def\@urlcolor{blue}%
   \def\@linkcolor{blue}%

\def\orcidID#1{\smash{\href{http://orcid.org/#1}{\protect\raisebox{-1.25pt}{\protect\includegraphics{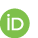}}}}}
\makeatother

\newcommand{\code}[1]{{\normalfont\texttt{#1}}}
\newcommand{\li}{\ \code{list}}
\newcommand{\tensor}{\code{tensor}}
\newcommand{\letin}[1]{\code{let}\ #1\ \code{in}\ }
\newcommand{\ifelse}[3]{\code{if}\ #1\ \code{then}\ #2\ \code{else}\ #3}
\newcommand{\fix}[1]{\code{fix}(#1)}
\newcommand{\blame}{\code{error}}
\newcommand{\red}[1]{{\color{red!100} #1}}

\renewcommand{\int}[0]{\code{int}}
\newcommand{\bool}[0]{\code{bool}}
\newcommand{\refine}[1]{\ensuremath{\Phi(#1)}}
\newcommand{\seq}[1]{\widetilde{#1}}
\newcommand{\p}{\vdash}
\newcommand{\assert}[1]{\mathbf{assert}(#1)}
\newcommand{\shape}[1]{#1.\code{shape}}
\newcommand{\len}[1]{\code{len}(#1)}
\newcommand{\nth}[1]{\code{nth}(#1)}

\newcommand{\reduce}{\longrightarrow}

\newcommand{\pred}{\varphi}
\newcommand{\ST}[1]{\code{BT}(#1)}
\newcommand{\dom}[1]{\mathit{dom}(#1)}
\newcommand{\typeof}[1]{\mathit{ty}(#1)}
\newcommand{\evof}[1]{\mathit{ev}(#1)}
\newcommand{\Iff}{\Leftrightarrow}
\newcommand{\wf}{\p_{\code{wf}}}
\newcommand{\st}{\p_{\code{BT}}}
\newcommand{\base}[1]{\{x:B\mid #1\}}
\newcommand{\COL}{\mathbin{:}}
\newcommand{\ocamltorch}{OCaml-Torch}
\newcommand{\tomono}{\Longrightarrow}

\newcommand{\BTE}[1]{\code{Base}(#1)}
\newcommand{\self}[1]{\textit{self}(#1)}
\renewcommand{\top}{\code{true}}
\newcommand{\graten}{\textsc{GraTen}}

\newenvironment{miniindent}{\qquad  \begin{minipage}{0.8\textwidth}}{\end{minipage}}

\usepackage{etoolbox}
\newcommand{\zerodisplayskips}{%
  \setlength{\abovedisplayskip}{3pt}%
  \setlength{\belowdisplayskip}{3pt}%
  \setlength{\abovedisplayshortskip}{2pt}%
  \setlength{\belowdisplayshortskip}{2pt}}
\appto{\normalsize}{\zerodisplayskips}
\appto{\small}{\zerodisplayskips}
\appto{\footnotesize}{\zerodisplayskips}

\let\emptyset\varnothing
\newif\ifdraft\drafttrue
\ifdraft
\newcommand\nk[1]{\red{[#1 -nk]}}
\newcommand\hattori[1]{\red{[#1 -hattori]}}
\else
\newcommand\nk[1]{}
\newcommand\hattori[1]{}
\fi

\newtheorem{assumption}{Assumption}

\begin{document}
\title{Gradual Tensor Shape Checking (full version)}
%
%\titlerunning{Abbreviated paper title}
% If the paper title is too long for the running head, you can set
% an abbreviated paper title here
%
\author{Momoko Hattori \and
Naoki Kobayashi\orcidID{0000-0002-0537-0604} \and
Ryosuke Sato\orcidID{0000-0001-8679-2747}}

\authorrunning{M. Hattori et al.}
% First names are abbreviated in the running head.
% If there are more than two authors, 'et al.' is used.

\institute{The University of Tokyo, Tokyo, Japan \\
\email{\{momohatt,koba,rsato\}@is.s.u-tokyo.ac.jp}}

\maketitle              % typeset the header of the contribution
\begin{abstract}
Tensor shape mismatch is a common source of bugs in deep learning programs.
We propose a new type-based approach to detect tensor shape mismatches.
One of the main features of our approach is the best-effort shape inference.
As the tensor shape inference problem is undecidable in general,
we allow static type/shape inference to be performed
only in a best-effort manner. If the static inference
cannot guarantee the absence of the shape
inconsistencies, dynamic checks are inserted into
the program. Another main feature is gradual typing, where users can improve
the precision of the inference by adding appropriate type annotations
to the program.
We formalize our approach and prove that it satisfies the criteria of gradual typing proposed by Siek et al. in 2015. We have implemented a prototype shape checking tool based on our approach and
evaluated its effectiveness by applying it to some deep neural network programs.

%\keywords{refinement types \and gradual typing.}
\end{abstract}

\section{Introduction}
\label{sec:intro}

\subsubsection*{Tensor Shape Checking and Its Difficulties.}

Tensor shape mismatch is one of the common sources of dynamic errors in programs using tensors (i.e., multi-dimensional arrays).
For example, the reshape operation of tensors takes a tensor $x$ and an integer list $S$ and returns a new tensor of the shape $S$ obtained by realigning the elements in $x$.
The input and output tensors must have the same number of elements;
a tensor of shape $[2; 3; 4]$\footnote{In this paper, we denote lists in the OCaml-style as in $[1; 2; 3]$ to disambiguate it from the citations.} can be reshaped into a shape $[3; 2; 4]$, while trying to reshape it into $[3; 4]$ results in a runtime error.

Early detection of tensor shape mismatch errors is critical in particular
for deep learning programs, where tensors are frequently used.
Since deep learning programs often take a considerable amount of time to train networks, it is often the case that a program takes hours and days to compute the weights of deep neural networks only to be terminated by one tensor shape mismatch error, throwing away the trained weights.
Even worse, some tensor shape mismatches can be harder to notice:
mixing up the height and the width of square images does not raise runtime errors but degrades the performance of the neural network.

The existing work on static detection of tensor shape mismatch errors
can be classified into two categories.
One is the whole-program analysis approach~\cite{pythia,shapeflow},
which collects tensor shape information
by partially evaluating the program in the style of abstract interpretation.
The other is the type-based approach~\cite{hasktorch,relay2018},
which expresses the shapes of tensors as a part of the type information.
Still, none of them is fully satisfactory: either they are too conservative and reject valid programs, or fail to detect some shape mismatch errors.

This paper pursuits the type-based approach as it is expected to provide modular detection of tensor shape inconsistencies.
Designing an appropriate type system and a type inference procedure to reason
about tensor shapes is challenging because shapes are first-class objects.
For example, the library function \texttt{Tensor.zeros} of
\ocamltorch{}~\cite{ocaml-torch} (which provides OCaml bindings for libtorch~\cite{pytorch})
takes a list $S$ of integers, and returns a new tensor whose shape is $S$. Thus, we have to work with \emph{dependent types}:
\texttt{Tensor.zeros} would be given the type
\( S\COL \code{int list} \to \{r:\tensor\mid \shape{r}=S\}\).
It is difficult to infer such dependent (refinement) types fully automatically.
Yet, we wish to avoid programmers' burden of writing too many type annotations.

Another difficulty is that shape constraints can be so complex
that even type checking, let alone inference, can be too costly or impossible.
For instance, the reshape operation explained earlier needs the proof that the shape of the input tensor $x$ is compatible with the given shape $S=[s_1;\ldots;s_n]$ (i.e., if the shape of $x$ is to be $[s_1';\ldots;s_m']$, then $\Pi_{i=1}^m s_i' = \Pi_{i=1}^n s_i$ holds)\footnote{Actually, some \(s_i\) can be \(-1\), in which case the size of the \(i\)-th dimension is unspecified.}.
Thus, type checking requires complex reasoning about (non-linear)
integer arithmetic and lists.

\subsubsection*{Overview of Our Approach.}

Based on the observations above, we propose an approach that
is expected to work well in practice despite the above-mentioned difficulties.
Our approach can be characterized by three main features: best-effort type inference, hybrid type checking, and gradual typing~\cite{siek2015refined}.
We explain them using our prototype tool \graten\footnote{The tool is publicly available at \url{https://doi.org/10.5281/zenodo.7590480}. The source code is also publicly available at \url{https://github.com/momohatt/graten}.}.

\paragraph{Best-Effort Type Inference.}

\graten{} does not try to infer the \emph{most general} types;
it performs type/shape inference in a \emph{best-effort} manner.
Thanks to this design choice, \graten{} works even if
no type annotations are provided (despite that the underlying type
system involves dependent types), and yet it can statically detect
(not necessarily all but) some shape mismatch errors.

\begin{figure}[bpt]
  %\qquad  \begin{minipage}{0.8\textwidth}
\begin{miniindent}
  \begin{Verbatim}
let model s =
  let f = ... in let g = ... in fun x -> let y = f x in g y
let _ = model 1 (Tensor.rand [20])
  \end{Verbatim}
  \end{miniindent}
  \vspace{5pt}
  \caption{An OCaml program written with OCaml-Torch.}
  \label{fig:example-intro}
\end{figure}

As an example, let us consider
the program in Figure~\ref{fig:example-intro}.
The function \code{model} takes an integer parameter \code{s}, defines functions \code{f} and \code{g}, and returns a layer
(which is a function that takes a tensor and returns a tensor)
which composes \code{f} and \code{g}.
The definitions of \code{f} and \code{g} are omitted here, but their types are
assumed as below, where \code{s} in the type of \code{f} is the argument of \code{model}
and the function $\nth{n,S}$ returns the $n$-th element of the list $S$ (the index
starts with 0).
\begin{align*}
  \code{f}:& \,\, x{:}\{ \nu:\tensor \mid \len{\shape{\nu}} = 1 \} \to \tensor\left(\left[\nth{0,\shape{x}}/\code{s}\right]\right) \\
  \code{g}:& \,\, \tensor([10]) \to \tensor([1])
\end{align*}
These types indicate that \code{f} takes a 1-dimensional tensor (i.e., a vector) and returns a vector whose length equals
the length of the argument vector divided by \code{s},
and that \code{g} expects a vector of length 10 and returns a vector of length 1.
The formal syntax of types will be introduced later in Section~\ref{sec:typing}.

For the program above,
\graten{}'s best-effort inference
outputs the following type for the function \code{model}.
{
  \footnotesize
\[
  s{:}\int
  \to x{:}\left\{ \nu{:}\tensor \mid \len{\shape{\nu}} = 1 \land \nth{0,\shape{\nu}}/\code{s} = 10 \right\}
  \to \tensor([1])
\]
}
Here, the constraint
${\nth{0,\shape{\nu}}}/{\code{s}}=10$
for the shape of $x$ is necessary for this program not to raise a shape mismatch error at the application of \code{g}.
The inferred type of \code{model} is used to prevent any calls to \code{model}
that violate the constraint.
Indeed, \graten{} rejects the call on line 4 of Figure~\ref{fig:example-intro},
where the arguments do not satisfy the constraint $\frac{\nth{0,\shape{\nu}}}{\code{s}} = 10$.
As in this example, our approach can statically detect shape mismatches
when enough type information has been obtained 
from the best-effort type inference or user-provided type annotations.

\paragraph{Hybrid Type Checking.}

\begin{figure}[bpt]
\begin{miniindent}
  \begin{Verbatim}
let model s =
  let f = ... in let g = ... in
  fun x -> let y = \textcolor{red!100}{if s = 1 then x else} f x in g y
  \end{Verbatim}
\end{miniindent}
  \vspace{5pt}
  \caption{The program from Figure~\ref{fig:example-intro} with small modification.}
  \label{fig:example-intro-2}
\end{figure}
\begin{figure}[bpt]
\begin{miniindent}
  \begin{Verbatim}
let model s =
  let f = ... in let g = ... in
  fun x -> let y = if s = 1 then x else f x in
           g (assert (y.shape = [10]); y)
  \end{Verbatim}
\end{miniindent}
  \vspace{5pt}
  \caption{The program returned by \graten{}
    given the program in Figure~\ref{fig:example-intro-2}.}
  \label{fig:example-cast-inserted-program}
\end{figure}

Another main feature of our approach is hybrid type checking:
we combine static and dynamic checking.
The type checker inserts assertions to
program points where the type safety is not statically guaranteed,
\`a la Knowles and Flanagan's hybrid type checking~\cite{hybridtyping}.
For example, consider the program in Figure~\ref{fig:example-intro-2}, which is obtained by adding a conditional branch to the one in Figure~\ref{fig:example-intro}.
The type of the \code{then} and \code{else} branch of the \code{if} expression
are inferred to be $\tensor(\shape{\code{x}})$ and $\tensor([\frac{\nth{0,\shape{\code{x}}}}{\code{s}}])$, respectively.
% Although these shapes in fact can be collectively expressed as $[\frac{\nth{0,\shape{\code{x}}}}{s}]$, the best-effort inference of \graten{} cannot come up with such a shape expression.
In this case, the type of \code{y} is simply inferred to be $\tensor$ without any information about its shape, and the inferred type for \code{model} is as follows.
\[
  s{:}\int \to x{:}\{ \nu:\tensor \mid \len{\shape{\nu}} = 1 \} \to \tensor([1])
\]
Thus, the best-effort inference of \graten{} fails
to capture the constraint $\frac{\nth{0,\shape{\nu}}}{s}=10$ for $x$ due to the imprecise type information of \code{y}.
Along with the inferred types, \graten{} outputs the program in Figure~\ref{fig:example-cast-inserted-program}, which is the same as the original program except for
the assertion inserted at the argument of \code{g}.
Since the statically inferred type of \code{y} fails to guarantee
that the application of \code{g} to \code{y} does not leads to a shape mismatch error,
\graten{} inserts the assertion to check the requirement dynamically.

\paragraph{Gradual Typing.}

\begin{figure}[bpt]
\begin{miniindent}
  \begin{Verbatim}
let model s =
  let f = ... in let g = ... in
  fun x ->
    let y = ((if s = 1 then x else f x) \textcolor{red!100}{: tensor([nth 0 x.shape / s])})
    in g y
  \end{Verbatim}
\end{miniindent}  
  \vspace{5pt}
  \caption{The program from Figure~\ref{fig:example-intro-2} after adding type annotations.}
  \label{fig:example-intro-3}
\end{figure}

Lastly, our approach incorporates \emph{gradual typing}~\cite{siek2015refined}\footnote{Usually, gradual typing introduces new syntax for gradual types and makes a distinction between static types and gradual types.
However, our type system does not have such distinction; it only uses the standard refinement types.
As we see later, we extend the standard refinement type system with cast (assertion) insertion rules so that it can be viewed as a gradualized type system.}
so that the users can improve the precision of
inferred types by adding type annotations.
For example, let us consider
the program in Figure~\ref{fig:example-intro-3}, which
is obtained from the one in
Figure~\ref{fig:example-intro-2} by adding a type annotation to \code{y}.
With this annotation, \graten{} infers the same type for \code{model} as it did
for \code{model} in Figure~\ref{fig:example-intro},
and no assertions are inserted.
As such, adding correct type annotations improves the type checking and decreases the number of assertions inserted.

Thanks to the best-effort inference, users need not
add type annotations to everywhere in the program.
They can focus on the program points where the static inference did not perform well, which is indicated by the insertion of assertions.
We prove
that our type system satisfies the gradual guarantee~\cite{siek2015refined}, which ensures that adding type annotation preserves the type-ability and the behavior of the program (with some assertions inserted) regardless of its precision, as long as the annotation does not disagree with the program.

\vspace{10pt}

Among the three features, the notion of hybrid type checking was first proposed by Knowles and Flanagan~\cite{hybridtyping}, and our gradual typing is closely related to gradual refinement types by \cite{lehmann2017gradual}, but we believe that the particular combination of three features is new.
In particular, unlike the original gradual refinement types~\cite{lehmann2017gradual}, we insert assertions instead of carrying around evidence terms~\cite{agt} in the reduction to guarantee type safety.

The contributions are summarized as follows.
\begin{inparaenum}[(i)]
  \item The formalization of a type system that combines hybrid type checking and gradual typing.
  We define our type system as the type-based transformation relation from source programs to programs with run-time assertion checks. We prove the soundness of our type system as well (Section~\ref{sec:typing}).
  The shape-polymorphic extension of the type system is also briefly discussed
  (Appendix~\ref{sec:poly}).
  \item A proof that our system satisfies the gradual guarantee~\cite{siek2015refined} (Section~\ref{sec:gradual}).
  \item Implementation of a best-effort type inference on a prototype system \graten{} inference (Section~\ref{sec:imp}).
  \item Experimental evaluation of \graten{} using the examples of deep learning programs bundled in the \ocamltorch{} library.
  We confirm that \graten{} can statically type-check the programs effectively with a reasonable amount of type annotations (Section~\ref{sec:experiment}).
\end{inparaenum}

\section{A Gradually-Typed Language with Refinement Types}
\label{sec:typing}

In this section, we formalize our type system and the translation to insert assertions.
We first introduce the source and target languages of the translation in
Sections~\ref{subsec:lang-term} and \ref{subsec:lang-cast-term}.
We then formalize the type system and the translation and prove their soundness in Section~\ref{subsec:typing-rules}.
The gradual guarantee is discussed later in Section~\ref{sec:gradual}.

\subsection{Source Language}
\label{subsec:lang-term}

We consider a call-by-value functional language,
whose syntax is given in Figure~\ref{fig:syntax}.
Throughout this paper, \(n\), \(c\), and \(x\) respectively denote integers, constants (including integers and primitive functions) and variables.
The base types $B$ and refinement predicates $\pred$ are explained later.

\begin{figure}[bpt]
  \begin{align*}
    M\,\text{(term)} &::= c \mid x \mid \lambda x{:}\tau.M \mid M \, x \mid (M:\tau)
      \mid \letin{x = M_1} M_2 \\
      &\mid\quad \fix{f{:}(x{:}\tau_1\to\tau_2),x,M}
      \mid \ifelse{x}{M_1}{M_2} \\
    \tau\,\text{(type)}
    &::= \{ x:B \mid \pred \}
    \mid x{:}\tau_1 \to \tau_2 \\
    \Gamma\,\text{(type env.)} &::= \emptyset \mid \Gamma,x:\tau  \qquad
    \Delta\,\text{(base type env.)}
    ::= \emptyset \mid \Delta,x:B
  \end{align*}
  \caption{Syntax of the source language, the types and the type environments.}
  \label{fig:syntax}
\end{figure}

Type annotations can be added to the function arguments $\lambda x{:}\tau.M$, recursive functions $\fix{f{:}(x{:}\tau_1\mathbin{\to}\tau_2),x,M}$ and to arbitrary expressions by $(M\mathbin{:}\tau)$.
In the implementation of \graten{}, users may omit the type annotations in lambda expressions and recursive functions as the best-effort type inference tries to complete them.

The argument of a function application and the branching condition of an if-expression are restricted to variables for the sake of simplicity of typing rules. Note that this restriction does not lose generality, as a general function application $M_1 \, M_2$ can be normalized to $\letin{f=M_1}\letin{x=M_2} f\,x$.

Types are defined following the standard definition of refinement types.
Intuitively, the type \(\{x\COL B\mid \pred\}\) describes a value \(x\) of type \(B\)
such that \(\pred\) holds. For example, \(\{x\COL \int\mid x \geq 0\}\) is the type
of non-negative ints.
We may omit the refinement predicates when they are \code{true}.
For example, we may write $\{ x\COL\int \mid \code{true} \}$ as $\int$.

The language presented so far is general;
in \graten{} it is instantiated to a language for tensor programs
by defining the base types and refinement predicates as in Figure~\ref{fig:predicate-graten},
and assuming that primitive operations on tensors are included in
the set of constants ranged over \(c\).
The refinement predicates, shapes and sizes are expressions of type $\bool$, $\int\li$ and $\int$ respectively.
The supported predicates are those described by quantifier-free formulas of first-order logic.
As shown in the definition, they may use some built-in predicates and functions over integer lists such as \code{append} and primitives on integer arithmetic in order to express common tensor operations.
We implicitly assume that the refinement predicates are well
formed (as defined in Figure~\ref{fig:well-formed-}, Appendix~\ref{sec:fulldef}).

\begin{figure}[bpt]
  \begin{align*}
    B\,\text{(base type)} &::= \bool \mid \int \mid \int\li \mid \tensor \\
    \pred\,\text{(predicate)} &::= \code{true} \mid \code{false} \mid s_1 = s_2 \mid S_1 = S_2 \mid x
    \mid \lnot \pred \mid \pred_1 \land \pred_2 \mid \pred_1 \lor \pred_2 \\
    &\mid\quad \code{broadcastable}(S_1,S_2) \mid \code{reshapeable}(S_1,S_2) \\
    S\,\text{(shape)} &::= [s_1;\ldots;s_n] \mid x \mid \shape{x} \mid \code{cons}(s,S) \mid \code{append}(S_1,S_2) \mid \code{tail}(S) \\
    &\mid\quad \code{init}(S) \mid \code{insertAt}(s_1,s_2,S) \mid \code{dropAt}(s, S) \mid \code{swap}(s_1,s_2,S) \\
    &\mid\quad \code{reshape}(S_1,S_2) \mid \code{broadcast}(S_1,S_2) \mid \code{matmul}(S_1,S_2) \\
    s\,\text{(size)} &::= n \mid x \mid -s \mid s_1 + s_2 \mid s_1\times s_2 \mid \frac{s_1}{s_2} \mid \code{head}(S) \mid \code{last}(S) \\
    &\mid\quad \len{S} \mid \nth{s, S} \mid \code{prod}(S)
  \end{align*}
  \caption{Syntax of base types $B$ and predicates $\pred$ in \graten{}.}
  \label{fig:predicate-graten}
\end{figure}

\subsection{Target Language}
\label{subsec:lang-cast-term}
\begin{figure}[tbp]
  \begin{align*}
    v\,\text{(value)} &::= c \mid x \mid [v_1,\ldots,v_n] \mid \lambda x^\tau. N \mid \fix{f^\tau, x, N} \\
    N\,\text{(cast term)} &::= v \mid \ifelse{v}{N_1}{N_2} \mid N\,v
       \mid \letin{x^{\tau} = N_1}N_2 \mid \assert{\pred}; N
  \end{align*}
  \caption{Syntax of the target language.}
  \label{fig:syntax-cast-term}
\end{figure}

As explained in Section~\ref{sec:intro}, we insert run-time checks into places where type-safety cannot be statically guaranteed.
Figure~\ref{fig:syntax-cast-term} shows the syntax of programs obtained by the insertion of assertions.
A main difference from the source language is the addition of assertion $\assert{\pred}; N$, which is used to implement the run-time checks.
Like Flanagan's hybrid type system~\cite{hybridtyping} (and unlike the blame calculus~\cite{wadler2009well}), we guarantee the safety of target programs by assertions.
Compared with the blame calculus, this method is expected to be easier to implement since most of the modern programming languages are equipped with assertions, and more efficient in that it avoids the accumulation of dynamic casts at runtime.
This implementation of the dynamic cast is possible since our system is only ``gradualized'' at the predicate level of the refinement type and the underlying simple type is static.

Another difference is that the binders in \code{let} expressions are annotated with their type.
This is required when defining the precision relation over the cast terms in Section~\ref{sec:gradual}.

\begin{figure}[tbp]
  \begin{minipage}[t]{0.6\textwidth}
    \begin{flushleft}
      \fbox{$[v/x]N$}
    \end{flushleft}
    \vspace{-10pt}
    \begin{alignat*}{2}
      [v/x](\assert{\pred}; N) &= \assert{[v/x]\pred}; [v/x]N \\
      [v/x](\lambda y^\tau.N) &= \lambda y^{[v/x]\tau}. [v/x] N
    \end{alignat*}
    (Variables are assumed to be alpha-renamed so that variables at different scopes do not collide)
  \end{minipage}
  \begin{minipage}[t]{0.35\textwidth}
    \begin{flushleft}
      \fbox{$N_1\reduce N_2$}
    \end{flushleft}
    \vspace{-10pt}
    \begin{alignat*}{2}
      \assert{\code{true}};N &\reduce N \\
      \assert{\code{false}};N &\reduce \blame \\
      c\,v &\reduce \code{ev}(c,v) \\
      (\lambda x^{\tau}.N_1)\,v &\reduce [v/x]N_1
    \end{alignat*}
  \end{minipage}

  \caption{Selected rules of substitution and reduction of the target language
    (the full definition is given
    in Figures~\ref{fig:reduction-} and ~\ref{fig:subst-type},
    Appendix~\ref{sec:fulldef}).}
  \label{fig:reduction}
\end{figure}

The substitution and the reduction rules of the cast terms are presented in Figure~\ref{fig:reduction}.
The evaluation of primitive function $\code{ev}(c, v)$ is defined to be the return value of the primitive function $c$ applied to an argument $v$ if $v$ meets the constraint of the argument of $c$, and otherwise undefined.
We denote $N \Uparrow$ if there exists an infinite reduction sequence from $N$.

The substitution for cast terms is defined in the standard manner, except that the implicitly-annotated type information and the predicate in the assertion need to be updated as well.
As can be seen in the definition of the cast term reduction, these implicitly-annotated types are only required for the sake of formalization and ignored at runtime.

\begin{figure}[tbp]
  \typicallabel{CT-Con}
  \begin{minipage}{0.31\hsize}
    \infax[CT-Con]{
      \Gamma;\pred\p c:\typeof{c}
    }
  \end{minipage}
  \begin{minipage}{0.32\hsize}
    \infrule[CT-VF]{
      \Gamma(x)=y{:}\tau_1\to\tau_2
    }{
      \Gamma;\pred\p x:\Gamma(x)
    }
  \end{minipage}
  \begin{minipage}{0.35\hsize}
    \infrule[CT-VB]{
      \Gamma(x)=\{y:B\mid\pred'\}
    }{
      \Gamma;\pred\p x:\{y:B\mid y=x\}
    }
  \end{minipage}

  \vspace{7pt}
  \begin{minipage}{0.44\hsize}
    \infrule[CT-Lam]{
      \Gamma,x:\tau_1;\pred\p N:\tau_2
    }{
      \Gamma;\pred\p \lambda x^{\tau_1}.N : x{:}\tau_1\to\tau_2
    }
  \end{minipage}
  \begin{minipage}{0.57\hsize}
  \typicallabel{CT-Const}
    \infrule[CT-App]{
      \Gamma;\pred\p N:x{:}\tau_1\to\tau_2
      \andalso \Gamma;\pred\p v:\tau_1
    }{
      \Gamma;\pred\p N\,v:[v/x]\tau_2
    }
  \end{minipage}

  \vspace{7pt}
  \begin{minipage}{0.58\hsize}
    \infrule[CT-Fix]{
      \Gamma,f:(x{:}\tau_1\to\tau_2),x:\tau_1;\pred\p N:\tau_2
    }{
      \Gamma;\pred\p \fix{f^{x{:}\tau_1\to\tau_2},x,N}:x{:}\tau_1\to\tau_2
    }
  \end{minipage}
  \begin{minipage}{0.41\hsize}
    \infrule[CT-Ass]{
      \Gamma;\pred\land\pred'\p N:\tau
    }{
      \Gamma;\pred\p \assert{\pred'};N : \tau
    }
  \end{minipage}

  \infrule[CT-If]{
    \Gamma;\pred\p v:\{x:\bool\mid\pred'\} \andalso
    \Gamma;\pred\land v\p N_1:\tau
    \andalso \Gamma;\pred\land\lnot v\p N_2:\tau
  }{
    \Gamma;\pred\p\ifelse{v}{N_1}{N_2}:\tau
  }

  \begin{minipage}[t]{0.48\hsize}
    \infrule[CT-Let]{
      \Gamma;\pred\p N_1:\tau_1
      \andalso \Gamma,x:\tau_1;\pred\p N_2:\tau
    }{
      \Gamma;\pred\p \letin{x^{\tau_1}=N_1}N_2 : \tau
    }
  \end{minipage}
  \begin{minipage}[t]{0.48\hsize}
  \typicallabel{CT-Const}
    \infrule[CT-Sub]{
      \Gamma;\pred\p N:\tau'
      \andalso \Gamma;\pred\p \tau'<:\tau
    }{
      \Gamma;\pred\p N:\tau
    }
  \end{minipage}
  \caption{Typing rules for the cast terms $\Gamma;\pred\p N:\tau$.}
  \label{fig:typing-rules-cast-term}
\end{figure}

\begin{figure}[tbph]
  \begin{minipage}[t]{0.44\hsize}
    \begin{flushleft}
      \fbox{$\refine{\Gamma}, \ST{\Gamma}$}
    \end{flushleft}
    \vspace{-15pt}
    \begin{align*}
      \refine{\emptyset} &= \code{true} \\
      \refine{\Gamma,x:\{y:B\mid\pred\}} &= \refine{\Gamma}\land[x/y]\pred \\
      \refine{\Gamma,x:(y{:}\tau_1\to\tau_2)} &= \refine{\Gamma} \\
      \ST{\emptyset} &= \emptyset \\
      \ST{\Gamma,x:\{x:B\mid\pred\}} &= \ST{\Gamma},x:B \\
      \ST{\Gamma,x:(y{:}\tau_1\to\tau_2)} &= \ST{\Gamma}
    \end{align*}
  \end{minipage}
  \begin{minipage}[t]{0.55\hsize}
    \begin{flushleft}
      \fbox{$\Gamma;\pred\p\tau_1<:\tau_2$}
    \end{flushleft}
    \infrule[Sub-Base]{
      \vDash\forall\ST{\Gamma},x{:}B.\refine{\Gamma}\land\pred\land\pred_1\Rightarrow\pred_2
    }{
      \Gamma;\pred\p \{x:B\mid\pred_1\} <: \{x:B\mid\pred_2\}
    }
    \infrule[Sub-Fun]{
      \Gamma;\pred\p \tau_3 <: \tau_1
      \andalso \Gamma,x:\tau_3;\pred\p\tau_2 <: \tau_4
    }{
      \Gamma;\pred\p x{:}\tau_1\to\tau_2 <: x{:}\tau_3\to\tau_4
    }
  \end{minipage}
  \caption{Subtyping rules.}
  \label{fig:subtype}
\end{figure}

We also introduce the type derivation rules for the cast terms $\Gamma;\pred\p N:\tau$ in Figure~\ref{fig:typing-rules-cast-term}.
This relation is used in the discussion of the soundness of the type system later in Section~\ref{subsubsec:type-safety}.
The quadruple relation $\Gamma;\pred\p N:\tau$ denotes that a cast term $N$ has type $\tau$ under a type environment $\Gamma$ and a logical context $\pred$.
The logical context $\pred$ holds the information of logically valid predicates at respective program points. New predicates are added at the \code{then} branch and the \code{else} branch of \rn{(CT-If)}, and the post-assertion cast term in \rn{(CT-Ass)}.
The subsumption is allowed in \rn{(CT-Sub)} by the subtyping relation $\Gamma;\pred\p\tau_1<:\tau_2$ (Figure~\ref{fig:subtype}), which is defined in a standard manner.

\subsection{Typing Rules}
\label{subsec:typing-rules}

\subsubsection{Inserting Assertions}
\begin{figure}[tbp]
  \vspace{-11pt}
  \begin{minipage}{0.495\hsize}
    \infax[CI-Const]{
      \Gamma;\pred\p c \leadsto c : \typeof{c}
    }
  \end{minipage}
  \begin{minipage}{0.495\hsize}
    \infrule[CI-Var-Fun]{
      \Gamma(x) = y{:}\tau_1\to\tau_2
    }{
      \Gamma;\pred\p x \leadsto x : \Gamma(x)
    }
  \end{minipage}

  \vspace{5pt}
  \begin{minipage}{0.495\hsize}
    \infrule[CI-Var-Base]{
      \Gamma(x) = \{y:B\mid\pred'\}
    }{
      \Gamma;\pred\p x \leadsto x : \{y:B\mid y=x\}
    }
  \end{minipage}
  \begin{minipage}{0.495\hsize}
    \infrule[CI-Lam]{
      \Gamma,x:\tau_1;\pred\p M \leadsto N : \tau_2
    }{
      \Gamma;\pred\p \lambda x{:}\tau_1.M \leadsto \lambda x^{\tau_1}.N : x{:}\tau_1\to\tau_2
    }
  \end{minipage}

  \infrule[CI-App]{
    \Gamma;\pred\p M_1\leadsto N_1:y{:}\tau_1\to\tau_2
    \andalso \Gamma(x)=\tau_3
    \andalso \Gamma;\pred\p \tau_3 \lesssim \tau_1 \leadsto N_2
  }{
    \Gamma;\pred\p M_1\,x \leadsto (\letin{x^{\tau_1}=N_2\,x}N_1\,x):[x/y]\tau_2
  }

  \infrule[CI-Fix]{
    \Gamma,f:(x{:}\tau_1\to\tau_2),x:\tau_1;\pred\p M \leadsto N :\tau_2
  }{
    \Gamma;\pred\p \fix{f{:}(x{:}\tau_1\to\tau_2),x,M}
    \leadsto \fix{f^{x{:}\tau_1\to\tau_2},x,N} : x{:}\tau_1\to\tau_2
  }

  \infrule[CI-Let]{
    \Gamma;\pred\p M_1 \leadsto N_1 : \tau_1
    \andalso \Gamma,x:\tau_1;\pred\p M_2 \leadsto N_2 : \tau
    \andalso \ST{\Gamma}\wf\tau
  }{
    \Gamma;\pred\p (\letin{x=M_1}M_2) \leadsto (\letin{x^{\tau_1}=N_1}N_2) : \tau
  }

  \infrule[CI-If]{
    \Gamma;\pred\p v : \{x:\bool\mid\pred'\}
    \andalso \Gamma;\pred\land v\p M_1 \leadsto N_1 : \tau
    \andalso \Gamma;\pred\land \lnot v\p M_2 \leadsto N_2 : \tau
  }{
    \Gamma;\pred\p \ifelse{v}{M_1}{M_2} \leadsto \ifelse{v}{N_1}{N_2} : \tau
  }

  \begin{minipage}{0.35\hsize}
    \infrule[CI-Annot]{
      \Gamma;\pred\p M \leadsto N : \tau
    }{
      \Gamma;\pred\p (M:\tau) \leadsto N : \tau
    }
  \end{minipage}
  \begin{minipage}{0.64\hsize}
    \infrule[CI-Sub]{
      \Gamma;\pred\p M_1 \leadsto N_1 : \tau_1
      \andalso \Gamma;\pred\p \tau_1 \lesssim \tau \leadsto N_2
    }{
      \Gamma;\pred\p M_1 \leadsto \letin{x^{\tau_1}=N_1}N_2\,x : \tau
    }
  \end{minipage}
  \caption{Type derivation rules for the source language $\Gamma;\pred\p M \leadsto N:\tau$.}
  \label{fig:typing-rules}
\end{figure}

Next, we discuss the typing rules for the source language and the assertion insertion into it.
Figure~\ref{fig:typing-rules} defines the type judgement and cast insertion relation.
The intuition of 5-ary relation $\Gamma;\pred\p M \leadsto N : \tau$ is: under a type environment $\Gamma$ and a logical context $\pred$, a term $M$ translates to a cast term $N$ and has type $\tau$.
If we ignore the part ``\(\leadsto N\)'' and replace the gradual subtyping relation
\(\lesssim\) with the standard subtyping relation on refinement types (Figure~\ref{fig:subtype}), our type
system is a standard refinement type system. Thus, the main novelty in
the rules in Figure~\ref{fig:typing-rules} lies in the use of the \emph{consistent subtyping relation}
$\Gamma;\pred\p\tau_1\lesssim\tau_2\leadsto N$, which is explained below.

The consistent subtyping relation $\Gamma;\pred\p\tau_1\lesssim\tau_2\leadsto N$ (Figure~\ref{fig:lesssim})\footnote{This can be understood as the refinement-type version of the differential subtyping in \cite{safe-typescript}, although in the implementation we do not calculate the ``difference'' between $\pred_1$ and $\pred_2$ for $\pred'$ in the assertion unless $\pred_1$ implies $\pred_2$ (and thus $\pred'$ can be \top).} is used in the cast insertion relation to guarantee that there exists a value that has both of the types $\tau_1$ and $\tau_2$ under $\Gamma$ and $\pred$, and to produce an assertion term $N$ that checks at runtime if a value that is statically known to be of type $\tau_1$ can be used as a value of type $\tau_2$.

The rule for the base case \rn{(Cast-Base)} checks if there exists a value, and an assignment of the values to the variables in the type environment, that satisfies both $\tau_1$ and $\tau_2$.
This intuitively holds if $\tau_1$ is castable to $\tau_2$ for some runtime values.
The rule also produces a lambda function that implements the cast with an assertion.
It is defined in such a way that $\pred_2$ can always be used as the content of the assertion $\pred'$, but $\top$ can also be used for $\pred'$ if $\pred_1$ implies $\pred_2$.
Note that we cannot use $\pred_2$ as the content of the assertion in the definition, or otherwise Proposition~\ref{prop:subtyping-lesssim} does not hold.

The rule for the function types \rn{(Cast-Fun)} recursively checks the castability of the argument types and the return types and combines the assertion terms for them.
Notice how the subsumption for the return types $\tau_2$ and $\tau_4$ has the meet of two argument types $\tau_1\sqcap\tau_3$ in the type environment.
The meet of two types (Figure~\ref{fig:lesssim}) is defined as a conjunction of the refinement predicates\footnote{Although the meet of two function types is defined, it does not make any difference in the definition of consistent subtyping relation since function types in the type environment is not used.}.

The consistent subtyping relation can be seen as a gradualization of the subtyping relation $\Gamma;\pred\p\tau_1<:\tau_2$ (Figure~\ref{fig:subtype}).
In fact, when a type $\tau_1$ is a subtype of another type $\tau_2$, it is possible that the assertion term generated by casting $\tau_1$ to $\tau_2$ only contains assertions that always succeed, which can be erased by some optimization.
The following proposition states this fact.
Note that this corresponds to the blame-subtyping theorem, one of the criteria for gradual typing presented in \cite{siek2015refined}.
\begin{figure}[tbp]
  \begin{align*}
    \{x:B\mid \pred_1\} \sqcap \{x:B\mid \pred_2\} &= \{x:B\mid \pred_1 \land \pred_2\} \\
    (x{:}\tau_1\to\tau_2) \sqcap (x{:}\tau_3\to\tau_4) &= x{:}(\tau_1\sqcap\tau_3) \to (\tau_2\sqcap\tau_4)
  \end{align*}

  \infrule[Cast-Base]{
    \vDash\exists\ST{\Gamma},x{:}B. \refine{\Gamma}\land\pred\land\pred_1\land\pred_2
    \andalso \vDash\forall\ST{\Gamma},x{:}B. \refine{\Gamma}\land\pred\land\pred_1\Rightarrow(\pred'\Iff\pred_2)
  }{
    \Gamma;\pred\p \{x:B\mid\pred_1\} \lesssim \{x:B\mid\pred_2\}
    \leadsto \lambda x^{\{x:B\mid\pred_1\}}.\assert{\pred'}; x
  }
  \infrule[Cast-Fun]{
    \Gamma;\pred\p \tau_3\lesssim\tau_1 \leadsto N_1
    \andalso \Gamma,x:\tau_1\sqcap\tau_3;\pred\p\tau_2\lesssim\tau_4 \leadsto N_2
  }{
    \Gamma;\pred\p x{:}\tau_1\to\tau_2 \lesssim x{:}\tau_3\to\tau_4
    \leadsto \\
    \lambda f^{x{:}\tau_1\to\tau_2}.\lambda x^{\tau_3}. (\letin{y^{\tau_1\sqcap\tau_3}=N_1\,x}\letin{z^{\tau_2}=f\,y}N_2\,z)
  }
  \vspace{-10pt}
  \caption{Definition of the consistent subtyping relation $\Gamma;\pred\p\tau_1\lesssim\tau_2\leadsto N$.}
  \label{fig:lesssim}
\end{figure}

\begin{proposition}\label{prop:subtyping-lesssim}
  $\Gamma;\pred\p \tau_1<:\tau_2$ implies $\Gamma;\pred\p\tau_1\lesssim\tau_2\leadsto N$ for some $N$ where all the assertions in $N$ are of the form $\assert{\code{true}}; N'$.
\end{proposition}

\subsubsection{Type Safety}
\label{subsubsec:type-safety}

We conclude this section with a note on the soundness of our type system.
The soundness is based on the fact that if the source program is well-typed, the program after the assertion insertion is also well-typed.

The most critical part of the proof is to prove the assertion term can be assigned a function type from the pre-assertion type to the post-assertion type.

\begin{lemma}\label{lemma:cast-term-type}
   $\Gamma;\pred\p\tau_1\lesssim\tau_2\leadsto N$ implies $\Gamma;\pred\p N:x{:}\tau_1\to\tau_2$ for some variable $x$ that does not occur in $\tau_2$.
\end{lemma}

This is an immediate consequence of Lemma \ref{lemma:cast-term-type-prep} in the appendix.
With Lemma~\ref{lemma:cast-term-type}, we can prove that the assertion-inserted program can be assigned the same type as that of the original program.

\begin{lemma}[Assertion Insertion Preserves Types]
  \label{lemma:translation-preserves-type}
  $\Gamma;\pred\p M\leadsto N:\tau$ implies $\Gamma;\pred\p N:\tau$.
\end{lemma}

We can also prove the standard progress and preservation properties
under a reasonable assumption that the types of the primitive functions are properly defined as follows
(see the appendix for the proofs).

\begin{assumption}\label{assume:evof}
  $\p c\,v:\tau$ implies $\evof{c,v}$ is defined and $\p \evof{c, v}:\tau$
\end{assumption}

Combining Lemma \ref{lemma:translation-preserves-type} with the progress and preservation properties, we obtain the type safety as follows.

\begin{theorem}[Type Safety]
  With Assumption \ref{assume:evof},
  $\emptyset;\code{true}\p M\leadsto N:\tau$ implies $N \reduce^* v$ for some $v$, $N \Uparrow$, or $N \reduce^* \blame$.
\end{theorem}

The type safety property states that a well-typed program does not cause untrapped dynamic errors.
The only case where a cast-inserted program causes untrapped errors is when the result of an application of a primitive function is undefined (i.e., $\evof{c,v}$ is undefined).
The type safety property ensures that such untrapped errors do not happen for well-typed terms as long as the $\typeof{c}$ is defined appropriately.

% \begin{remark}
%   In our system, we used refinement types to express tensor shapes.
%   An alternative design would be to express the type of a tensor with shape $S$ as $\tensor(S)$ (instead of $\{x:\tensor\mid\shape{x}=S\}$) and incorporate shape polymorphism and dynamic shapes/sizes.
%   In such a system, the type of ReLU function would be $\forall S.\tensor(S) \to \tensor(S)$ and the type of a 1-dimensional tensor with statically unknown length would be $\tensor([\star])$, where $\star$ denotes the unknown size.
%   Although this approach may look simpler, it introduces an unnecessary complication.

%   Consider a function of type $\forall a, b. \tensor([a \times b]) \to \tensor([a + b])$, and applying it to a 1-dimensional tensor with statically unknown length $\tensor([\star])$.
%   The polymorphic shape parameters $a$ and $b$ would both need to be instantiated with $\star$, since no other option would be more appropriate.
%   Then, we would need to consider a dynamic cast from $\tensor([\star])$ to $\tensor([\star\times\star])$, which would require an existential quantifier in the assertion.

%   Since we observed through the experiments that the types of most of the realistic tensor functions can be expressed using our refinement types, we did not choose this alternative design.
%   Some exceptional cases we have noticed are discussed in the next section.
% \end{remark}

\end{document}

\section{Gradual Guarantee}
\label{sec:gradual}

In a standard gradual type system, programs are compared by their \emph{precision}, or the amount of information contained in the type annotations.
This notion is used to define the gradual guarantee~\cite{siek2015refined}, which is the core property of gradual typing.
The gradual guarantee comes in two parts. The first one is called \emph{static gradual guarantee}, which states that decreasing the precision of type annotation from a well-typed program still preserves the typeability of the program at a less precise type.
The second one is called \emph{dynamic gradual guarantee}, which claims that a less precise program behaves the same as the more precise one with fewer assertion errors.

Below we first define the precision for the language introduced in Section~\ref{sec:typing}.
We then show that our type system satisfies the gradual guarantee.

\subsubsection*{Precision.}
%\label{subsec:precision}

\begin{figure}[tbp]
  \begin{minipage}[t]{0.55\hsize}
    \begin{flushleft}
      \fbox{$\seq{x}\p\tau_1\sqsubseteq\tau_2$}
    \end{flushleft}
    \vspace{-20pt}
    \infrule[Prec-Base]{
      \vDash\forall\seq{y},x.\pred_1\Rightarrow\pred_2
    }{
      \seq{y}\p \{x:B\mid\pred_1\} \sqsubseteq \{x:B\mid\pred_2\}
    }
    \infrule[Prec-Fun]{
      \seq{y}\p \tau_1 \sqsubseteq \tau_3
      \andalso \seq{y},x\p\tau_2 \sqsubseteq \tau_4
    }{
      \seq{y}\p x{:}\tau_1\to\tau_2 \sqsubseteq x{:}\tau_3\to\tau_4
    }
  \end{minipage}
  \begin{minipage}[t]{0.44\hsize}
    \begin{flushleft}
      \fbox{$\Gamma_1\sqsubseteq\Gamma_2$}
    \end{flushleft}
    \vspace{-20pt}
    \infax{
      \emptyset\sqsubseteq\emptyset
    }
    \infrule{
      \Gamma_1 \sqsubseteq \Gamma_2
      \andalso \dom{\Gamma_1}\p\tau_1\sqsubseteq\tau_2
    }{
      \Gamma_1,x:\tau_1 \sqsubseteq \Gamma_2,x:\tau_2
    }
  \end{minipage}
  \caption{Precision relation of types and type environments.}
  \label{fig:precision}
\end{figure}
Figure~\ref{fig:precision} defines the precision relation $\seq{x}\p\tau_1\sqsubseteq\tau_2$ on types by using the logical implication between the refinement predicates.
The sequence of variables $\seq{x}$ keeps the variables that may appear in the refinement predicates.
For example, the following is an example of the type precision relation for the base type.
\[
  \p \{x:\tensor\mid\shape{x}=[3]\} \sqsubseteq \{x:\tensor\mid\len{\shape{x}}=1\} \\
\]

Note that in the rule \rn{(Prec-Fun)}, the precision of the argument type and the return type are compared independently;
the type information on $x$ is not used in the comparison of the return types. This is in contrast with the rule \rn{(Sub-Fun)} in Figure~\ref{fig:subtype} for subtyping.
Figure~\ref{fig:precision} also extends the relation to $\Gamma\sqsubseteq\Gamma'$ on type environments.
The precision relation is also extended to the relation $\seq{x}\p M \sqsubseteq M'$ on terms, by the rules in Figure~\ref{fig:precision-term}. Here, $\seq{x}$ is the sequence of variables in scope.
Finally, we define the precision relation of the cast terms in Figure~\ref{fig:precision-term}.
Unlike the term precision relation (Figure~\ref{fig:precision-term}), the precision relation $\Gamma;\pred\p N_1\sqsubseteq N_2$ on cast terms requires the type environment $\Gamma$ and the logical context $\pred$ in the judgement, and the refinement extraction from the type environment $\refine{\Gamma}$ is used in the rule \rn{(PC-Assert)}.
We also assume the following property on the evaluation of the primitive functions.
\begin{assumption}
  If $\evof{c,v_2}$ and $\evof{c,v_1}$ are both defined, then $v_1\sqsubseteq v_2$ implies $\evof{c,v_1}\sqsubseteq\evof{c,v_2}$
\end{assumption}

\begin{figure}[tbp]
  \begin{flushleft}
    \fbox{$\seq{x}\p M_1 \sqsubseteq M_2$}
  \end{flushleft}
  \centering
  \begin{minipage}{0.45\hsize}
    \infrule[PM-Lam]{
      \seq{y}\p \tau_1\sqsubseteq\tau_2
      \andalso \seq{y},x\p M_1 \sqsubseteq M_2
    }{
      \seq{y}\p \lambda x{:}\tau_1.M_1 \sqsubseteq \lambda x{:}\tau_2.M_2
    }
  \end{minipage}
  \begin{minipage}{0.5\hsize}
    \infrule[PM-Annot]{
      \seq{y}\p M_1 \sqsubseteq M_2
      \andalso \seq{y}\p \tau_1 \sqsubseteq \tau_2
    }{
      \seq{y}\p (M_1 : \tau_1) \sqsubseteq (M_2 : \tau_2)
    }
  \end{minipage}
  \begin{flushleft}
    \fbox{$\Gamma;\pred\p N_1\sqsubseteq N_2$}
  \end{flushleft}
  \vspace{-10pt}
  \infrule[PC-Assert]{
    \forall\ST{\Gamma}.\refine{\Gamma}\land\pred\land\pred_1\Rightarrow\pred_2
    \andalso \Gamma;\pred\land\pred_1\p N_1 \sqsubseteq N_2
  }{
    \Gamma;\pred\p \assert{\pred_1};N_1 \sqsubseteq \assert{\pred_2};N_2
  }
  \caption{Selected rules for the precision relation on terms and cast terms (the full definition is found in the appendix Figure~\ref{fig:precision-term-} and Figure~\ref{fig:precision-cast-term-} respectively).}
  \label{fig:precision-term}
\end{figure}

Intuitively, the precision of cast terms are designed in such a way that, when $\emptyset;\code{true}\p N_1\sqsubseteq N_2$ holds, the assertions in $N_1$ is more strict than that of $N_2$, and therefore the dynamic checks in $N_1$ is more likely to fail than in $N_2$.
The following two propositions state this intuition (the proofs are found
in the appendix).

\begin{proposition}\label{prop:reduction-left}
  Suppose $\emptyset;\top\p N_1:\tau$ and $\emptyset;\top\p N_2:\tau'$.
  Then, $\emptyset;\code{true}\p N_1\sqsubseteq N_2$ and $N_1\reduce N_1'$ imply $N_2\reduce N_2'$ and $\emptyset;\code{true}\p N_1'\sqsubseteq N_2'$ for some $N_2'$.
\end{proposition}

\begin{proposition}\label{prop:reduction-right}
  Suppose $\emptyset;\top\p N_1:\tau$ and $\emptyset;\top\p N_2:\tau'$.
  Then, $\emptyset;\code{true}\p N_1\sqsubseteq N_2$ and $N_2\reduce N_2'$ imply either of the following.
  \begin{itemize}
    \item $N_1\reduce N_1'$ and $N_1'\sqsubseteq N_2'$ for some $N_1'$
    \item $N_1\reduce \blame$
  \end{itemize}
\end{proposition}

\subsubsection{Gradual Guarantee.}
%\label{subsec:gradual-guarantee}
We show that our system satisfies the gradual guarantee~\cite{siek2015refined}.
First, we prove that the consistent subtyping relation $\Gamma;\pred\p\tau_1\lesssim\tau_2\leadsto N$ is upper-closed with respect to the precision relation $\seq{x}\p\tau_1\sqsubseteq\tau_3$ on types.

\begin{lemma}\label{lemma:lesssim-relax}
  $\Gamma;\pred\p\tau_1\lesssim\tau_2\leadsto N_1$,
  $\dom{\Gamma}\p\tau_1\sqsubseteq\tau_3$,
  $\dom{\Gamma}\p\tau_2\sqsubseteq\tau_4$ and
  $\Gamma\sqsubseteq\Gamma'$ implies
  $\Gamma';\pred\p\tau_3\lesssim\tau_4\leadsto N_2$ for some $N_2$.
\end{lemma}

We can further prove that the cast term $N_2$ in the statement of Lemma~\ref{lemma:lesssim-relax} is less precise than the original cast term $N_1$ as follows.

\begin{lemma}\label{lemma:cast-precision}
  Suppose $\Gamma\sqsubseteq\Gamma', \dom{\Gamma}\p\tau_1\sqsubseteq\tau_1'$ and $\dom{\Gamma}\p\tau_2\sqsubseteq\tau_2'$.
  Then, $\Gamma;\pred\p\tau_1\lesssim\tau_2\leadsto N$ and $\Gamma';\pred\p\tau_1'\lesssim\tau_2'\leadsto N'$ implies $\Gamma;\pred\p N\sqsubseteq N'$.
\end{lemma}

Using the above properties, we can prove the following lemma which constitutes the core part of the proof of the gradual guarantee.

\begin{lemma}\label{lemma:static-gradual-guarantee}
  $\Gamma\sqsubseteq\Gamma'$, $\dom{\Gamma}\p M \sqsubseteq M'$ and $\Gamma;\pred\p M \leadsto N : \tau$ imply
  $\Gamma';\pred\p M' \leadsto N' : \tau'$, $\Gamma;\pred\p N \sqsubseteq N'$ and $\dom{\Gamma}\p\tau\sqsubseteq\tau'$ for some $N'$ and $\tau'$.
\end{lemma}

Finally, we can show the static and dynamic gradual guarantee as follows.

\begin{theorem}[Static gradual guarantee]
  $\emptyset\p M_1 \sqsubseteq M_2$ and $\p M_1 :\tau_1$ imply
  $\p M_2 :\tau_2$ and $\emptyset\p \tau_1 \sqsubseteq \tau_2$ for some $\tau_2$.
\end{theorem}

\begin{proof}
 This follows immediately from Lemma~\ref{lemma:static-gradual-guarantee}. \qed
\end{proof}

\begin{theorem}[Dynamic gradual guarantee]
  Suppose $\emptyset\p M_1 \sqsubseteq M_2$ and $\p M_1 \leadsto N_1:\tau_1$.
  Then, there exist $N_2$ and $\tau_2$ that satisfy all of the following.
  \begin{itemize}
    \item $\p M_2 \leadsto N_2 : \tau_2$.
    \item $N_1\reduce^*v_1$ implies $N_2\reduce^*v_2$ and $v_1\sqsubseteq v_2$ for some $v_2$.
    \item $N_1\Uparrow$ implies $N_2\Uparrow$.
    \item $N_2\reduce^*v_2$ implies $N_1\reduce^*v_1$ and $v_1\sqsubseteq v_2$ for some $v_1$, or $N_1\reduce^*\blame$.
    \item $N_2\Uparrow$ implies $N_1\Uparrow$ or $N_1\reduce^*\blame$.
  \end{itemize}
\end{theorem}

\begin{proof}
  By Lemma~\ref{lemma:static-gradual-guarantee}, $\p M_2\leadsto N_2:\tau_2$ holds for some $N_2$ and $\tau_2$ where $\p N_1 \sqsubseteq N_2$ and $\p \tau_1 \sqsubseteq \tau_2$.
  Also, from Lemma~\ref{lemma:translation-preserves-type}, we obtain $\p N_1:\tau_1$ and $\p N_2:\tau_2$.
  Using Proposition~\ref{prop:reduction-left}, $N_1 \reduce^* v_1$ for some $v_1$ implies $N_2 \reduce^* v_2$ for some $v_2$ such that $v_1 \sqsubseteq v_2$.
  Also, $N_1 \reduce^\infty$ implies $N_2 \reduce^\infty$.
  Using Proposition~\ref{prop:reduction-right}, $N_2 \reduce^* v_2$ for some $v_2$ implies $N_1 \reduce^* v_1$ for some $v_1$ such that $v_1 \sqsubseteq v_2$, or $N_1 \reduce^* \blame$.
  Also, $N_2 \reduce^\infty$ implies $N_1 \reduce^\infty$ or $N_1 \reduce^* \blame$. \qed
\end{proof}

\section{Best-Effort Type Inference}
\label{sec:imp}

Thanks to our combination of gradual typing and hybrid checking described in
the previous sections, a type inference procedure need not necessarily output
the most precise types. It is allowed to perform type inference only in a
\emph{best-effort} manner, and
the results in the previous sections
do not depend on the particular design of the type inference procedure.
Nevertheless, it is desirable for the procedure to infer reasonably good types.
In this section, we report a specific design of the type inference procedure,
which we have implemented in our prototype system \graten{}; as reported
in the Section~\ref{sec:experiment}, our procedure works reasonably well
for actual deep learning programs.

\subsection{Overview of Type Inference and Checking in \graten{}}

The type checking in \graten{} consists of the following three phases:
(1) simple type inference,
(2) best-effort refinement type inference, and
(3) consistent subtyping checking and assertion insertion.

In the first phase, \graten{} performs the simple type inference using the standard Hindley-Milner algorithm and annotates the AST with the inferred simple types of each node.

In the second phase, \graten{} first collects all the consistent subtyping constraints of the form $\Gamma;\pred\p\tau_1\lesssim\tau_2\leadsto N$
from the source program.
When it encounters AST nodes whose refinement type cannot be constructed directly, \graten{} generates template refinement types using the simple types inferred in the previous phase.
Template refinement types may
contain variables for undetermined predicates (referred to as \emph{predicate variables}).

Using the collected constraints, \graten{} then tries to find a solution for all of the predicate variables with its hand-made constraint solver.
The constraint solving takes place on every \code{let} binding to allow let-polymorphism on shapes.
We discuss the detail of the implementation of the solver in the next subsection,
but at a high level, the solver tries to find such a solution that:
\begin{itemize}
  \item only general types are inferred, as otherwise it could result in rejecting well-typed programs.
  \item $\Gamma;\pred\p \tau_1<:\tau_2$ holds for as many constraints
  $\Gamma;\pred\p\tau_1\lesssim\tau_2\leadsto N$ as possible.
  This is to make the cast term \(N\) consist of trivial assertions (which
  can statically be discharged to avoid run-time overheads;
  recall Proposition~\ref{prop:subtyping-lesssim}).
\end{itemize}
Given that the subtyping constraints can be expressed in the form of constrained Horn clauses (CHC) and not all the subtyping constraints need to hold, the problem above is essentially a CHC solving problem
with weak constraints and maximality~\cite{prabhu2021specification} where the optimization objective of the problem is defined by pointwise logical comparison of the solutions.

The constraint solver of 
\graten{} does not always find a solution for all predicate variables.
In such cases, \graten{} assigns \code{true} to the undetermined predicate variables; that way, they will at least not invalidate the consistent subtyping constraints.

Note that \graten{} does not take into account the consistent subtyping $\Gamma;\pred\p\tau_1\lesssim\tau_2\leadsto N$ itself when trying to find a solution,
as we expect that it would be rare for a consistent subtyping $\Gamma;\pred\p\tau_1\lesssim\tau_2\leadsto N$ to hold when the subtyping relation $\Gamma;\pred\p\tau_1<:\tau_2$ does not hold.
\graten{} therefore defers the check of consistent subtyping constraints to the next phase.

In the third phase, \graten{} checks the validity of consistent subtyping constraints using the inference results for the predicate variables from the previous phase. \graten{} first attempts to simplify and verify the constraints by a hand-made solver, but it falls back on using z3~\cite{z3} with timeouts if it does not work. Simultaneously, it also generates the assertion terms and inserts them into the source program.

\subsection{Heuristics of Best-Effort Type Inference}

To solve the subtyping constraints
explained above,
we have implemented a hand-made constraint solver.
\graten{} does not use off-the-shelf SMT or CHC solvers such as Z3~\cite{z3},
since the refinement predicates in \graten{} often use complicated predicates on integer lists,
for which standard SMT/CHC solvers cannot find a solution in a reasonable time.
Also, while \graten{} should infer general types (so as not to reject well-typed programs),
those generic solvers are not biased towards generality and return any (non-general)
solution that satisfies the constraints.
This subsection describes the heuristics used in \graten{}
for constraint solving.

The preparation for the inference is already started when \graten{} generates the template refinement types during the constraint collection.
For each predicate variable generated, \graten{} attaches the set of program variables it depends on, which is calculated from the type environment.
This is used in the constraint solving later to avoid assigning irrelevant predicates to
the predicate variables.
We denote predicate variables as $p_{\seq{x}}(\seq{y})$, where $\seq{x}$ denotes the set of program variables it depends on and $\seq{y}$ denotes the parameters of the predicate variable.

After collecting the constraints, \graten{} decomposes the subtyping constraints to
constrained Horn clauses of the form $\seq{\pred_1}\land\seq{\pred_2}\Rightarrow\seq{\pred_3}$
following the definition of the subtyping relation (Figure~\ref{fig:subtype}).
The notation $\seq{\pred}$ denotes a set of predicates,
logically interpreted as the conjunction of the predicates.
The first, second, and third set of predicates in the clause respectively corresponds to the predicates from the context $\refine{\Gamma}\land\pred$, the refinement of the type on the left $\pred_1$, and that of the type on the right $\pred_2$. We intentionally distinguish between $\seq{\pred_1}$ and $\seq{\pred_2}$ on the left-hand side of the clauses in describing the constraint solving algorithm.
For example, let us reconsider the program in Figure~\ref{fig:example-intro-2}.
The subtyping constraints collected from the \code{if} expression of the program would be as follows, where $p, q$ and $r$ are the predicate variables generated for the type of \code{s}, \code{x} and the \code{if} expression respectively.
\begin{align*}
  \Gamma;(s=1)&\p \{\nu{:}\tensor\mid q_{s,\nu}(\nu)\} <: \{\nu{:}\tensor\mid r_{s,x,\nu}(\nu)\} \\
  \Gamma;(s\neq1)&\p \{\nu{:}\tensor\mid q_{s,\nu}(\nu)\} <: \{\nu{:}\tensor\mid \len{\shape{\nu}}=1\} \\
  \Gamma;(s\neq1)&\p \tensor([\nth{0,\shape{x}}/s]) <: \{\nu{:}\tensor\mid r_{s,x,\nu}(\nu)\} \\
  \text{where }\, \Gamma &:= [s\mapsto\{\nu{:}\int\mid p_{\nu}(\nu)\},x\mapsto\{\nu{:}\tensor\mid q_{s,\nu}(\nu)\}]
\end{align*}
These constraints are decomposed into the following clauses.
{
  \footnotesize
\begin{equation}
\begin{aligned}
  \{p_s(s), q_{s,x}(x), s=1\}\land \{ q_{s,\nu}(\nu) \} &\Rightarrow r_{s,x,\nu}(\nu) \\
  \{p_s(s), q_{s,x}(x), s\neq1\}\land \{ q_{s,\nu}(\nu) \} &\Rightarrow \len{\shape{\nu}} = 1 \\
  \{p_s(s), q_{s,x}(x), s\neq1\}\land \{ \shape{\nu}=[\nth{0,\shape{x}}/s] \} &\Rightarrow r_{s,x,\nu}(\nu)
\end{aligned}\label{eq:clause1}
\end{equation}
}

From the clauses obtained as above, \graten{} tries to find a solution for the predicate variables using an algorithm presented in Algorithm~\ref{alg:inference-heuristics}.

The algorithm processes the constraints by first trying to find a solution for
predicate variables that occur on the right-hand side of a clause $\seq{\pred_1}\land\seq{\pred_2}\Rightarrow\seq{\pred_3}$ (Line 6-10), and then on the left-hand side of a clause (Line 11-15),
and repeats it until either all of the constraints are solved or the constraints cannot be processed any further (Line 4).
In Line 8 and Line 13, the set of program variables $\seq{x}$ of a predicate variable $p_{\seq{x}}$ is used to assign the predicates to the predicate variables\footnote{The set of program variables used in predicates is defined following the standard definition of free variables, except that the program variables used in a predicate variable $p_{\seq{x}}$ is
defined as $\seq{x}$.}.

During the iteration, the constraints need to be occasionally updated with the current solutions $\theta$ by applying the substitution $\theta$ to all the predicates in the constraints. After that, we also simplify the set of clauses (with \code{simplify} in Algorithm~\ref{alg:inference-heuristics}) by removing the predicates from the right-hand side of a clause that trivially follows from the left-hand side, and by removing clauses whose right-hand side is empty.
For example, a clause $\{\}\land\{x=1\}\Rightarrow\{x=1\}$ is simplified to $\{\}\land\{x=1\}\Rightarrow\{\}$, and then removed from the set of clauses.

To illustrate the behavior of Algorithm~\ref{alg:inference-heuristics}, consider applying it to the clauses \eqref{eq:clause1}.
During the first iteration of the \code{while} loop (Line 4), the first \code{for} loop (Line 6) exits with an empty $\theta$ as $r$ appears on the right-hand side of multiple clauses and cannot be resolved here due to the check at Line 7.
In the next \code{for} loop (Line 11), $\theta$ is updated to:
\begin{equation}
  [q_{s,\nu}(\nu)\mapsto\left(\len{\shape{\nu}}=1 \land q'_{s,\nu}(\nu)\right)] \label{eq:solution}
\end{equation}
where $q'_{s,\nu}(\nu)$ is a fresh predicate variable, and the constraints $c$ would be updated as follows.
{
  \footnotesize
\begin{align*}
  \{p_s(s), \len{\shape{x}}=1, q'_{s,x}(x), s=1\}\land \{ \len{\shape{\nu}}=1 \land q'_{s,\nu}(\nu) \} &\Rightarrow r_{s,x,\nu}(\nu) \\
  \{p_s(s), \len{\shape{x}}=1, q'_{s,x}(x), s\neq1\}\land \{ \shape{\nu}=[\nth{0,\shape{x}}/s] \} &\Rightarrow r_{s,x,\nu}(\nu)
\end{align*}
}
The \code{while} loop exits after the second iteration, as no new predicate variables can be added to $\theta$ and $c=c'$ holds.
Thus, we only obtain \eqref{eq:solution} from Algorithm~\ref{alg:inference-heuristics}.
After the inference, \graten{} assigns \code{true} to the remaining predicate variables $p$, $q'$ and $r$.

\begin{algorithm}
  \hspace*{\algorithmicindent} \textbf{Input:} constrained Horn clauses $c$ \\
  \hspace*{\algorithmicindent} \textbf{Output:} the mapping from predicate variables to its solution (predicates) $\theta$
  \begin{algorithmic}[1]
    \Procedure{solve}{$c$}
    \State Let $\theta$ be an empty substitution
    \State $c' \gets c$
    \While{$c\neq \emptyset$ and $c \neq c'$}
      \State $c' \gets c$
      \For{every clause of the form $\seq{\pred_1}\land\seq{\pred_2}\Rightarrow p_{\seq{x}}(\seq{y})$ in $c$}
        \If{$p_{\seq{x}}(\seq{y})\notin\seq{\pred_3}'$ for any other $\seq{\pred_1}'\land\seq{\pred_2}'\Rightarrow\seq{\pred_3}'$ in $c$}
          \State Let $\seq{\pred_2}'$ be the maximal subset of $\seq{\pred_2}$ that only uses variables in $\seq{x}$
          \State $\theta \gets [p_{\seq{x}}(\seq{y})\mapsto\bigwedge\seq{\pred_2}'] \circ \theta$
          \Comment{$\circ$ is a composition of mappings.}
        \EndIf
      \EndFor
      \State $c \gets \code{simplify}(\theta \, c)$
      \Comment{$\code{simplify}(\cdot)$ is described in the main text.}
      \For{$\seq{\pred_1}\land\seq{\pred_2}\Rightarrow\seq{\pred_3}$ in $c$}
        \For{every predicate variable $p_{\seq{x}}(\seq{y})$ in $\seq{\pred_1}\cup\seq{\pred_2}$}
          \State Let $\seq{\pred_3}'$ be the maximal subset of $\seq{\pred_3}$ that only uses variables in $\seq{x}$
          \State Let $q_{\seq{x}}(\seq{y})$ be a fresh predicate variable
          \State $\theta \gets [p_{\seq{x}}(\seq{y})\mapsto\left(\bigwedge\seq{\pred_3}'\right)\land q_{\seq{x}}(\seq{y})] \circ \theta$
        \EndFor
        \State $c \gets \code{simplify}(\theta \, c)$
        \Comment{Also updates the remaining items iterated by L11.}
      \EndFor
    \EndWhile
    \State \Return $\theta$
    \EndProcedure
  \end{algorithmic}
  \caption{Algorithm for calculating the solutions $\theta$ to predicate variables from constrained Horn clauses $c$.}
  \label{alg:inference-heuristics}
\end{algorithm}

\section{Experiment}
\label{sec:experiment}

This section reports on experiments to evaluate the effectiveness of our approach
by running our tool \graten{} for the example programs bundled in the \ocamltorch{} library~\cite{ocaml-torch}. We have also checked how type annotations changed the inference results.

\subsection{Methods}

\subsubsection{Input and Output of \graten{}}

\graten{} takes an OCaml program and performs type checking with its best-effort type inference.
If the type checking is successful, it returns the inferred types of top-level variables defined in the program, and the source program with necessary assertions inserted.
Otherwise, the type checking fails with an error message.

The assertions are inserted into the output program only when they are needed.
Namely, assertions are inserted into the places where the consistent subtyping $\Gamma;\pred\p\tau_1\lesssim\tau_2\leadsto N$ is used only when $\Gamma;\pred\p\tau_1<:\tau_2$ doesn't hold (see Proposition~\ref{prop:subtyping-lesssim}).

Besides the source program, \graten{} also reads the types of the library functions (including those of OCaml-Torch) from manually prepared stub files.
For example, the type of \code{tr} (matrix transpose function) is defined as follows.
\begin{verbatim}
val tr :  x:{ v:tensor | len v.shape = 2 }
       -> tensor([nth 1 x.shape; nth 0 x.shape])
\end{verbatim}
Note that describing the types of some higher-order OCaml-Torch functions requires the polymorphic extension, which we sketch in
Appendix~\ref{sec:poly}.
For example, the type of \code{Layer.forward} is defined as follows.
\begin{align*}
  &\forall b_1{:}\bool,b_2{:}\bool.\\
  &(x{:}\{x{:}\tensor\mid b_1\}\to\{y{:}\tensor\mid b_2\}) \to x{:}\{x{:}\tensor\mid b_1\}\to\{y{:}\tensor\mid b_2\}
\end{align*}
\graten{} handles such types by instantiating the quantified parameters ($b_1$ and $b_2$ in the above case) with fresh predicate variables.

\subsubsection{Test Cases}
\label{subsubsection:test-cases}

We applied \graten{} to programs under \code{examples/} directory of the repository of \ocamltorch\footnote{\url{https://github.com/LaurentMazare/ocaml-torch/tree/a6499811f4/examples}}.
The list of programs tested is shown in Table~\ref{table:result}.
Since some programs use features of OCaml or \ocamltorch{} that are not yet supported by \graten{}, they were modified not to use such features without changing the structure of the neural network.
Major modifications added to the target programs are listed below.
Other smaller syntactic modifications can be found in the supplementary materials.

\begin{enumerate}[label=(M\arabic*)]
  \item \label{modif:poly} Replacing or removing type-polymorphic functions.
  Some functions that create loops such as \code{List.foldl} are replaced with recursive functions.
  Others such as \code{no\_grad} are replaced with the type-instantiated versions.
  \item \label{modif:list} Removing use of non-integer lists, especially tensor lists and layer\footnote{Functions that take a tensor and return a tensor.} lists. As a result, two list-taking primitive functions are removed.
  One is \code{Tensor.cat}, which takes a list of tensors and returns the concatenation of them. It is replaced with a variant \code{Tensor.cat\_} which takes only two tensors.
  The other is \code{Layer.sequential}, which takes a list of layers and returns a layer that sequentially applies all the input layers.
  \item \label{modif:ref-float} Replacing mutable float objects with 0-dimensional tensors, as \graten{} does not support reference types.
\end{enumerate}

As an example of \ref{modif:poly} and \ref{modif:list}, consider the following function,
which creates a list of linear layers and returns a new layer that applies all the layers in the list.
\begin{Verbatim}[fontsize=\small]
let f vs ~num_layers =
 List.init num_layers ~f:(fun i -> Layer.linear vs ~input_dim:(i+1) (i+2))
 |> Layer.sequential
\end{Verbatim}
The \code{i}-th layer in the list takes a tensor whose last dimension is size \code{i+1}, and returns a tensor of the same shape except that the last dimension is changed to \code{i+2}.
By the modifications \ref{modif:poly} and \ref{modif:list},
the above function definition is replaced with:
\begin{Verbatim}[fontsize=\small]
let f vs ~num_layers =
 let rec loop i xs =
  if i = 0
   then Layer.id xs
   else loop (i-1) xs ~is_training |> Layer.linear vs ~input_dim:i (i+1)
 in Layer.of_fn (loop num_layers)
\end{Verbatim}

Some programs in the \code{examples/} directory are excluded from the test cases for the following reasons.
\begin{asparaitem}
  \item \code{neural\_transfer} uses a library function \code{Vgg.vgg16\_layers} whose type cannot be described in \graten; the relation between its inputs and its output tensor's shape could not be expressed in the syntax supported by \graten{}.
  \item Programs \code{dqn.ml}, \code{dqn\_atari.ml} and \code{dqn\_pong.ml} in \code{reinforcement-learning} use queues which are not supported in \graten{} yet.
  \item \code{env\_gym\_pyml.ml} and \code{venv\_env\_gym\_pyml.ml} under \code{reinforcement-learning} use Python objects whose verification is not the scope of this paper.
  \item \code{reinforcement-learning/policy\_gradient.ml} uses mutable lists which cannot be replaced with another datatype already supported in \graten.
  \item \code{yolo/darknet.ml} and \code{translation/lang.ml} use hash tables which are not supported in \graten{} yet.
  \item \code{translation/dataset.ml} and \code{translation/lang.ml} are irrelevant as tensor objects do not appear in them.
\end{asparaitem}

\subsubsection{Evaluation}

We evaluated the best-effort inference of \graten{} on the following three aspects.

First, we counted the assertions inserted into the original program when \graten{} is used for the target program.
Since the assertions indicate the program points that could fail at runtime, the user of \graten{} would
wish to pay attention to the location and the number of inserted assertions and try to decrease them.

Second, we counted the minimum number of type annotations required to type-check the program with minimum assertions inserted. This is for evaluating
the realistic programmers' burden of trying to statically verify the program with type annotations.
The annotations were added in such a way that the types of the functions do not lose the original generality.
The type annotations are counted by the number of refinement types with non-\code{true} refinement predicates in them.
For example, the following annotation counts as 3
because the refinement of the input tensor and the two output tensors are not \code{true}, but the refinement of the annotation of the second argument \code{bool} is \code{true}.
\begin{verbatim}
tensor([x]) -> ~is_training:bool -> tensor([x]) * tensor([x])
\end{verbatim}

Third, we also measured the time taken by \graten{} to analyze the unannotated and annotated programs.
The experiments were conducted on a Linux machine with 12-core Intel i5-11400 (2.60GHz)
and \graten{} is implemented in Haskell with GHC version 9.0.2.

\subsection{Experimental Results}

\begin{table*}
  \begin{tabular}{|l|r||c|c||c|c|c|} \hline
  \multirow{2}{*}{Location under examples/} & \multirow{2}{*}{LOC} & \multicolumn{2}{c||}{Unannotated} & \multicolumn{3}{c|}{Annotated} \\ \cline{3-7}
                                    &  & time (s) & \#assert & \#annot & time (s) & \#assert \\ \hline\hline
  char\_rnn/char\_rnn.ml            &  98 & 1.647 &  1 &  2 & 0.664 & 0 \\ \hline
  cifar/cifar\_train.ml             &  72 & 0.311 &  0 &  - & - & - \\ \hline
  cifar/densenet.ml                 & 116 & 2.603 &  6 &  2 & 1.304 & 0 \\ \hline
  cifar/fast\_resnet.ml             &  64 & 0.293 &  0 &  - & - & - \\ \hline
  cifar/preact\_resnet.ml           &  85 & 2.535 &  8 &  5 & 0.346 & 0 \\ \hline
  cifar/resnet.ml                   &  78 & 2.597 &  8 &  4 & 0.396 & 0 \\ \hline
  gan/began.ml                      & 220 & 1.581 &  1 &  - & - & - \\ \hline
  gan/gan\_stability.ml             & 224 & 4.441 & 40 &  2 & 1.410 & 2 \\ \hline
  gan/mnist\_cgan.ml                & 117 & 0.498 &  1 &  - & - & - \\ \hline
  gan/mnist\_dcgan.ml               & 136 & 1.418 &  4 &  2 & 0.500 & 0 \\ \hline
  gan/mnist\_gan.ml                 &  83 & 0.308 &  0 &  - & - & - \\ \hline
  gan/progressive\_growing\_gan.ml  & 118 & 0.734 &  0 &  - & - & - \\ \hline
  gan/relativistic\_dcgan.ml        & 171 & 0.659 &  1 &  - & - & - \\ \hline
  jit/load\_and\_run.ml             &  16 & 0.214 &  1 &  - & - & - \\ \hline
  min-gpt/mingpt.ml                 & 207 & 3.036 &  8 &  6 & 2.686 & 0 \\ \hline
  mnist/conv.ml                     &  53 & 0.250 &  0 &  - & - & - \\ \hline
  mnist/linear.ml                   &  50 & 0.235 &  0 &  - & - & - \\ \hline
  mnist/nn.ml                       &  39 & 0.210 &  0 &  - & - & - \\ \hline
  pretrained/finetuning.ml          &  69 & 0.294 &  0 &  - & - & - \\ \hline
  pretrained/predict.ml             &  68 & 0.303 &  2 &  - & - & - \\ \hline
  reinforcement-learning/a2c.ml     & 105 & 0.418 &  0 &  - & - & - \\ \hline
  reinforcement-learning/ppo.ml     & 129 & 0.438 &  0 &  - & - & - \\ \hline
  reinforcement-learning/rollout.ml &  91 & 0.734 &  9 &  5 & 0.425 & 1 \\ \hline
  translation/seq2seq.ml            & 258 & 3.800 & 11 & 34 & 1.023 & 3 \\ \hline
  vae/vae.ml                        &  78 & 1.233 &  4 & 10 & 0.312 & 0 \\ \hline
  yolo/yolo.ml                      & 144 & 1.027 &  4 &  1 & 0.985 & 3 \\ \hline
  \end{tabular}
  \caption{Results of running \graten{} to the test cases.
  The second column is the size of the program after the modification.
  The third and fourth columns are the results for unannotated programs.
  The third column is the duration of the type-checking and the fourth column is the number of assertions inserted.
  From the fifth to the seventh columns are for the annotated programs. The fifth column is the number of annotations added to the program.
  }
  \label{table:result}
\end{table*}

Table~\ref{table:result} summarizes the experimental results.
We analyze those results by the following three aspects: assertions, type annotations and analysis time.

\subsubsection{Inserted Assertions}

Out of the 26 programs tested, 10 programs required no type annotations to type-check without assertions,
and other 7 programs type-checked without assertions
after adding appropriate type annotations.
For the remaining 9 programs such as gan/began.ml and gan/gan\_stability.ml,
we could not eliminate all assertions, although some of them were removed after adding type annotations.
The remaining assertions were due to the imprecise type signatures of some library functions.
For instance, \code{Torch.Serialize.load} is a function that loads a tensor from a file and its type signature is defined as follows.
\begin{verbatim}
val load : ~filename:string -> tensor
\end{verbatim}
The return type of \code{load} is simply defined as \code{tensor} since it is impossible to assume any properties about its shape.
As a result, an assertion was inserted to check if the loaded tensor satisfies the requirement to run the program without uncaught errors.
Even adding type annotations to the loaded tensor does not remove the assertion.

Some other functions are given imprecise types due to \graten{}'s immature support of polymorphic data types.
For example, the type of \code{Tensor.stack} is defined as follows because \graten{} does not effectively support non-integer lists yet. Refining the return types of such functions is left as future work.
\begin{verbatim}
val stack : ~dim:int -> list (tensor) -> tensor
\end{verbatim}

\subsubsection{Patterns of Added Type Annotations}

As we added type annotations to the test cases, we observed that the program points that require type annotations have similarities.
All of the type annotations fall into one of the following patterns.
\begin{enumerate}[label=(P\arabic*)]
  \item \label{pat:branch} Branches i.e., \code{if} expressions and \code{match} expressions with multiple branches (e.g., Figure~\ref{fig:example-intro-3} in Section~\ref{sec:intro}).
  \item \label{pat:recursion} Recursive functions. For example, \code{loop} in translation/seq2seq.ml is annotated as follows.
\begin{verbatim}
let rec loop
  :  ~state:tensor([1; enc.hidden_size])
  -> ~prevs:list ({ v:tensor | prod v.shape = 1 })
  -> ~max_length:int -> list ({ v:tensor | prod v.shape = 1 })
= fun ~state ~prevs ~max_length -> ...
\end{verbatim}
  \item \label{pat:ho} Higher-order shape-polymorphic arguments.
  For example, \code{sample} in \code{char\_rnn.ml} is annotated as follows.
\begin{verbatim}
let sample ~dataset ~lstm
  ~linear:(linear : x:{ v:tensor | last v.shape = hidden_size }
                 -> tensor(init x.shape @ [dataset.labels]))
  ~device = ...
\end{verbatim}
  \item \label{pat:record} Definition of record types. The current implementation of \graten{} expects that the definition of record types describes the refinement types of each field.
  \item \label{pat:primitive} Imprecise type signatures of primitive functions, or user-defined functions of dependent modules.
  For example, translation/seq2seq.ml has the following type annotation since the return type of \code{Tensor.stack} is only inferred to be \code{tensor} due to its imprecise type signature.
\begin{verbatim}
let enc_outputs : tensor([1; nth 1 v.shape; enc.hidden_size]) =
  Tensor.stack enc_outputs ~dim:1
\end{verbatim}
  The statically inferred type of \code{enc\_outputs} here is \code{tensor([1; enc.hidden\_size]) list}, so we would not need this type annotation if the type signature of \code{Tensor.stack} is appropriately defined.
  Since it is not possible to statically verify the correctness of these types of annotations, assertions would still be inserted after adding these annotations.
\end{enumerate}
The first three patterns indicate that \graten{}'s current
best-effort type inference does not effectively infer precise refinements for branches, recursive functions and higher-order shape-polymorphic arguments.
The fourth pattern \ref{pat:record} would be inevitable when using record types.
It remains as future work to exempt users from having to add type annotations for \ref{pat:primitive}.
With such improvements, we believe that it will become easier
to find program points that require type annotations for better inference.

\subsubsection{Number of Type Annotations}

There is no correlation between the number of assertions inserted into the unannotated program and the number of annotations needed to the program to minimize the number of assertions.

For example, adding two type annotations to \code{gan/gan\_stability.ml} resulted in removing 38 assertions. This is because \graten{} inferred an imprecise type for a helper function \code{resnet\_block} without any type annotations, and it degraded the precision of the inference for the 24 callers of the function.
Meanwhile, \code{translation/seq2seq.ml} required comparatively many type annotations as it has many definition of record types and several recursive functions with multiple inputs.

\subsubsection{Analysis Time}

For all of the 11 annotated programs,
\graten{}'s type checking for annotated programs was faster than
the unannotated counterparts.
This would be because having more static information made it easier for \graten{}
to infer more precise types and resolve more subsumption constraints easily.

\subsection{Discussions}
\label{subsec:discussions}

In this subsection, we discuss the strengths, weaknesses and our perspective on the future development of our system.

\subsubsection{Performance of Best-Effort Inference}

As reported in the previous subsection,
the best-effort inference of \graten{} does not infer precise types for branches, recursions and higher-order shape-polymorphic arguments.
While this may seem unsatisfying at a glance, the aim of this research is not to develop a perfect inference algorithm, but to propose a method that can work on unannotated programs and allows users to work interactively with the type checker to gradually add type annotations.
With this respect, we believe that \graten{} has achieved desirable results
since it will be easy for the user to find out where to add type annotations.
This is because (1) the inserted assertions can inform the user of the location of potential dynamic errors, and (2) all of the required type annotations would fall into one of the patterns listed in the previous section and thus should be predictable.

\subsubsection{Lists of Tensors and Layers}

As of now, the refinement inference for lists in \graten{} is limited to integer lists.
Meanwhile, lists of tensors or lists of functions are commonly used in deep learning programs:
\code{Tensor.cat} and \code{Tensor.stack} both take a list of tensors and return their concatenation,
and \code{Layer.sequential} takes a list of layers (functions that take and return a tensor) and returns their composition.

A potential approach to support these library functions would be to add new refinement predicates for tensors lists or layer lists.
For example, we can add a predicate $\code{composable}(x,S_1,S_2)$ which means that the composition of a list of layers $x$ takes a tensor of shape $S_1$ and returns a tensor of shape $S_2$.
The type of \code{Layer.sequential} would be expressed with the shape polymorphic extension
(see Appendix~\ref{sec:poly}) as follows.
\begin{verbatim}
val sequential : forall S1 S2.
  { v:list(tensor -> tensor) | composable(x,S1,S2) }
                                  -> tensor(S1) -> tensor(S2)
\end{verbatim}
To practically infer \code{composable} predicate for layer lists, we would need to change the type-instantiated versions of list-manipulating functions as well.
For instance, the type of the \code{cons} function for layers would need to be defined as follows.
\begin{verbatim}
val cons_layers
  :  forall S1 S2 S3. (tensor(S1) -> tensor(S2))
  -> { v:list(tensor -> tensor) | composable(v, S2, S3) }
  -> { v:list(tensor -> tensor) | composable(v, S1, S3) }
\end{verbatim}

\subsubsection{Reporting Incorrect Type Annotations}

Since our type system sees the standard refinement types as gradual,
some users might find the behavior of \graten{} unexpected in some cases.
Consider the following function \code{f} which takes a matrix and returns a matrix obtained by transposing the input.
Suppose that the programmer mistakenly annotated the return value of \code{f} to have the same shape as the input matrix.
\begin{verbatim}
let f x = (tr x : tensor(x.shape))
\end{verbatim}
Although this type annotation does not hold in general, this program is not rejected by our type system beause the annotation can hold if the input \code{x} is a square matrix.
\graten{} would output the following program with an assertion.
\begin{verbatim}
let f x = (fun y -> assert(y.shape = x.shape); y) (tr x)
\end{verbatim}
To avoid such a situation, it would be possible to extend the type system with
types with fully statically known refinements,
and let the annotated types be interpreted as such.

\section{Related Work}
\label{sec:rel}

\subsubsection{Tensor Shape Checking in Deep Learning Programs.}
The problem of tensor shape checking has been studied for decades by various contexts
such as the numeric analysis~\cite{eaton2006statically,abe2015simple} and the array-oriented languages with rank polymorphism~\cite{slepak2014array,slepak2018rank,gibbons2016aplicative}.
Tensor shape checking for deep learning programs is still a new challenge because the shapes can be more complicated, and a variety of methods have been proposed both in academia and in industry.

Some tools statically check tensor shapes with advanced type systems.
Hasktorch~\cite{hasktorch} is a Haskell binding of libtorch~\cite{pytorch} which provides a mode that statically checks tensor shapes.
Since they use the type-level programming feature of Haskell to implement the tensor shapes, tensor shapes are not first-class objects.
As a result, programs such as the one in Figure~\ref{fig:example-intro} cannot be expressed since it is impossible to define the function \code{f} whose type depends on the first-class object \code{s}.
Relay~\cite{relay2018,relay2019} is an IR for deep learning compilers with a rich type system for tensor shape with type inference.
Both Relay and Hasktorch support dynamic shape as a wild card in the static shape checking.

% Some earlier studies on the type-based approach~\cite{typesafe-scala,tensorsafe} also exist.
% A work by Chen~\cite{typesafe-scala} demonstrates an implementation of static shape checking in Scala by using type parameters and phantom types. Their method, however, would be difficult to extend for rank-polymorphic types or complex shape expressions.
% A deep learning library TensorSafe~\cite{tensorsafe} also uses the type-level programming of Haskell to statically check tensor shape, but their method does not verify shape-polymorphic functions on its own.

Apart from the type-based verification methods, some tensor shape error detection tools also take a static approach.
Pythia~\cite{pythia,ariadne} statically detects shape fault for TensorFlow~\cite{tensorflow} programs by keeping track of the tensor shapes throughout the program using value-flow analysis.
The tracking of shape is in a best-effort manner, allowing the shape inference results to be ``unknown'' in some cases.
The analysis crucially relies on the programming practice in TensorFlow to annotate tensor shapes as much as possible.

Other static checking tools took an approach that uses symbolic execution to collect constraints from the program and verifies it with a solver; Tensors Fitting Perfectly~\cite{tensors-fitting-perfectly} and PyTea~\cite{pytea} are on this approach.
Both methods remove loops from the program in an ad-hoc manner based on a reasonable assumption for the program.

Lastly, some took dynamic approaches to provide lightweight shape fault detection.
ShapeFlow~\cite{shapeflow} is an abstract interpreter of TensorFlow programs; it shares the same APIs as TensorFlow but only calculates the shape of tensors.
Users can run the analysis by replacing the import of TensorFlow with ShapeFlow in the target program, which executes more efficiently than the original TensorFlow program.
Elichika~\cite{elichika} uses a similar method to ShapeFlow with a feature to display the interpreted shapes with a symbolic expression.
These dynamic approaches enable quick analysis and require no type annotations, but provide no guarantee for untested inputs.

\subsubsection{Static and Dynamic Checking for Refinement Types.}
Earlier work on dependent type system focused on decidable type checking and inference with restricted refinement logic~\cite{refinementML,dml,xi1998eliminating,liquid}.
%Meanwhile,
Dynamic checking with contracts~\cite{eiffel,findler2002contracts} offers expressive verification that cannot be covered with a static type system, but at a cost of runtime overhead.
Naturally, the combination of static and dynamic checking has been actively explored by the successors of both parties.

Hybrid type checking~\cite{hybridtyping}, which our work is based on, extends the purely-dynamic method of using contracts by verifying specifications statically as much as possible.
This method differs from ours in that it inserts a dynamic check only when the subtyping constraint is not proven to be valid or invalid.
As a result, this method statically rejects the incorrectly annotated program that we discussed in Subsection~\ref{subsec:discussions}, while our method accepts it with a dynamic check in the hope that a more precise type annotation will remove the need for a dynamic check.
Our method can be understood as a variant of hybrid type checking with a focus on being gradual in adding type annotations.

% An earlier work by Ou et al.~\cite{ou2004dynamic} is similar to hybrid type checking, but differs in requiring the programmer to specify which part of the program needs to type check with refinement types and which part of the program does not need to.
% The programmer does not have to annotate the program segments marked as the latter, and dynamic checks are inserted as needed.

The application of gradual typing to dependent type systems has also been studied~\cite{lehmann2017gradual,eremondi2019approximate}.
Especially, gradual refinement types~\cite{lehmann2017gradual} is very similar to our type system in that it gradualizes only the predicate part of a refinement type system and the underlying simple type is static.
One of the differences is that their system distinguishes statically-unknown refinement predicates with statically-known ones, while our system assumes that any refinement predicates can have a statically-unknown portion.
For example, consider the following program:
\[
  \code{let}\,\,f\,x\,(y:\{\nu:\int\mid\top\}) = x / y
\]
This program is rejected in their system because the type annotation of $y$ indicates that the programmer is confident that $y$ can be any integers including 0; otherwise, the type annotation should have been $\{\nu:\int\mid\,\star\,\}$.
Meanwhile, our system interprets the type annotation as not precise enough and accepts the program by inserting a dynamic check to $y$.
Intuitively, $\{x:B\mid\pred\}$ in our type system translates to $\{x:B\mid\pred\,\land\, \star\}$ in gradual refinement types~\cite{lehmann2017gradual}.

The type inference for gradual refinement types has been studied by Vazou et al.~\cite{vazou2018gradual}.
Their work restricts the refinement to liquid predicates~\cite{liquid} to maintain the decidability, while our work does not impose such a limitation.

\section{Conclusion and Future Work}
\label{sec:conc}
We presented an extension to the standard refinement type system which
can be viewed as a gradual type system.
The essence of this extension is the introduction of the consistent subtyping relation, which inserts to the source program assertions that checks statically-unverified properties at runtime.
We also presented that the extended type system satisfies the refined criteria of gradual typing.

We then applied this type system for verifying tensor shapes with best-effort type inference.
This application makes use of the property of the proposed type system that allows us to cover the limitation of the static best-effort analysis with dynamic checks.
We also implemented a prototype type checker \graten{} and applied it with some of the example programs publicly available in \ocamltorch{} repository.
We observed that, thanks to the best-effort type inference, users would not be required too many type annotations to statically type-check the whole program, and it would not be difficult to find where to add type annotations to improve the inference.

We conclude with some ideas for future work.
\begin{asparaitem}
  \item Extension with type polymorphism.
  As we observed in the experiments, type polymorphic functions are frequently used in realistic programs.
  Extending our type system with ML-style type polymorphism would make the type checker more practical.

  \item Application for imperative languages with a dynamic type system, like Python. In this paper, we have chosen OCaml as the target of the prototype to ensure that the input program is statically-typed. Python would, however, be a more attractive target
    since it is widely used in the machine learning community.
\end{asparaitem}

\subsubsection*{Acknowledgments}
  We would like to thank anonymous referees for useful comments.
This work was supported by
JSPS KAKENHI Grant Number JP20H05703.
% \subsection*{Acknowledgment}
% This work was partially supported by JSPS KAKENHI Grant Numbers JP20H05703.

\bibliographystyle{splncs04}
\bibliography{main}

\appendix

\section{Extension of Type System with Polymorphism}
\label{sec:poly}

Using the refinement types, we can express the types of most of the shape polymorphic tensor functions.
For example, ReLU is a function that takes a tensor and returns a tensor of the same shape as the input.
In a type system with shape polymorphism, the type of such function is expressed as $\forall S{:}\int\li.\,\,\tensor(S) \to \tensor(S)$,
whereas in our system, it is expressed as $x{:}\tensor \to \tensor(\shape{x})$.

There is, however, shape polymorphism that cannot be expressed in this style.
Consider the type:
\[
  \forall S_1,S_2.\,\,(\tensor(S_1)\to\tensor(S_2))\to\tensor(S_1)\to\tensor(S_2)
\]
The application function $\lambda f.\lambda x. f\,x$, for example, has this type.
In our system presented so far, there is no way to present the equivalent of this type.

We therefore discuss an extension of our type system with explicit polymorphism on sizes and shapes.
We introduce type schemes and redefine type environments to
map from variables to type schemes.
\begin{align*}
    T\,\text{(type scheme)} &::= \forall x_1{:}B_1,\ldots,x_n{:}B_n. \tau \\
    \Gamma\,\text{(type environment)} &::= \emptyset \mid \Gamma[x\mapsto T]
\end{align*}
We define the new typing relation as a transformation relation
$\Gamma;\Delta;\pred\p M:\tau\tomono M'$
which associates a polymorphically-typed program $M$ with a monomorphically-typed program $M'$:
\infrule[PT-Var]{
  \Gamma(f)=\forall x_1{:}B_1,\ldots,x_n{:}B_n. \tau
  \andalso \ST{\Gamma},\Delta\st t_i:B_i\mbox{ for each $i$}
}{
  \Gamma;\Delta;\pred\p f: [t_1/x_1,\ldots,t_n/x_n]\tau \tomono f\,t_1\,\cdots\,t_n
}
\infrule[PT-Let]{
  \Gamma;(\Delta,x_1:B_1,\ldots,x_n:B_n);\pred\p M_1:\tau_1\tomono M_1'
  \andalso \mbox{\(x_1,\ldots,x_n\) do not occur in \(p\)} \\
  (\Gamma,f:\forall x_1{:}B_1,\ldots,x_n{:}B_n.\tau_1);\Delta;\pred\p M_2:\tau_2 \tomono M_2'
}{
  \Gamma;\Delta;\pred \p (\letin{f=M_1}{M_2}):\tau'
  \tomono(\letin{f=\lambda x_1{:}B_1.\cdots\lambda x_n{:}B_n.M_1'}{M_2'})
}
Here, $\Delta$ is a type environment for polymorphic shape variables.

Thus, a type scheme \(\forall x_1{:}B_1,\ldots,x_n{:}B_n. \tau\)
is mapped to a monomorphic (refinement) type
\(x_1{:}B_1\to\cdots\to x_n{:}B_n\to \tau\).
The term obtained by type inference is thus just a
monomorphically-typed term, to which the theory of
gradual tensor types developed in the previous subsections apply.

There is, however, a subtle conflict in the best-effort inference of
type schemes and the coercion mechanism of gradual typing.
In the rule \rn{(PT-Let)}, we need to infer appropriate
\(t_1,\ldots,t_n\) and choosing wrong \(t_1,\ldots,t_n\)
may cause cast failures, even if the original program is safe.
For example, consider the following program:
\begin{verbatim}
let app f x = f x in
let g = ... (* complex function on tensors *) in
  ... app g ...
\end{verbatim}
Then, the function \texttt{app} can be assigned the following
type scheme.
\[
  \forall S_1,S_2:\int\li.\,
  (\tensor(S_1)\to \tensor(S_2))\to \tensor(S_1)\to\tensor(S_2)
\]
Type inference then tries to convert the caller \(\code{app}\,\code{g}\,x\)
%and \(f\,h\) to terms of the form
to a term of the form \(\code{app}\,t_1\,t_2\,\code{g}\,x\).
%%and \(f\,t_1'\,t_2'\,h\).
However, if \texttt{g} is a complex function for which the precise
inference of shapes is difficult, then we will fail
to infer appropriate parameters \(t_1,t_2\).
In that case, we need to fall back to the safe side, and
assign to \texttt{app} a less precise monomorphic type:
\begin{align*}
(\tensor\to\tensor)\to\tensor\to\tensor.
\end{align*}
The caller \(\code{app}\,\code{g}\) is then just transformed to \(\code{app}\,\code{g}\,x\).

The solution above (of falling back to monomorphic typing when
type inference for callers fail) still has the following problems.
\begin{enumerate}
\item In the example above,
  other callers of \texttt{app} will also be monomorphically typed, which leads
  to imprecise type inference. For example, suppose \texttt{h} is statically known
  to have type \(\tensor([2;3])\to \tensor([1])\). Based on the type scheme,
  we could infer the type of \(\code{app}\,\code{h}\) to be
  \(\tensor([2;3])\to\tensor([1])\), but due to the presence of the caller
  \(\code{app}\,\code{g}\), the imprecise type
  \(\tensor\to \tensor\) is inferred also for \(\code{app}\,\code{h}\).
\item It is against the principle of modular type inference that
  whether a polymorphic type is assigned to \texttt{app} depends on callers.
\end{enumerate}
A remedy to the problems above is to prepare both
polymorphic and monomorphic versions for each polymorphic function.
In the example above, thus \code{app} is transformed to
two functions:
\begin{verbatim}
let app_poly s1 s2 f x = ...
let app_mono f x = ...
\end{verbatim}
where:
\begin{align*}
&\code{app\_poly}:\forall S_1,S_2.\,(\tensor(S_1)\to \tensor(S_2))\to \tensor(S_1)\to\tensor(S_2)\\
&\code{app\_mono}: (\tensor\to \tensor)\to \tensor\to\tensor
\end{align*}
Then we can transform \(\code{app}\,\code{g}\) and \(\code{app}\,\code{h}\)
to \(\code{app\_mono}\,\code{g}\) and \(\code{app\_poly}\, [2;3]\,[1]\,\code{h}\).

\begin{remark}
  The polymorphic extension we have discussed here may still not be suitable for the inference of some programs.
  For example, let $f$ be a function of the following type:
  \[
    f: \{x:\tensor\mid\code{prod}\,\,\shape{x} = 128\}\to\tensor
  \]
  and consider applying the above-mentioned \code{app} to $f$.
  Since it is impossible to infer the shape of the argument or the return value of $f$, the type of \code{app} falls back to that of \code{app\_mono}.
  The type of $\code{app}\,f$ is thus inferred to be $\tensor\to\tensor$,
  although we know that $\code{app}\,f$ can be assigned the same type as $f$.

  Function \code{Layer.forward} of \ocamltorch{} is one of the most frequently used primitive function, and it has similar types as \code{app}.
  We noticed that not being able to infer the same type as $f$ for $\code{app}\,f$ in the above case critically degrades the precision of inference.
  Therefore, in the prototype implementation, we specially assign the following refinement-polymorphic type to \code{Layer.forward}:
  \begin{align*}
    &\forall b_1{:}\bool,b_2{:}\bool.\\
    &(x{:}\{x{:}\tensor\mid b_1\}\to\{y{:}\tensor\mid b_2\}) \to x{:}\{x{:}\tensor\mid b_1\}\to\{y{:}\tensor\mid b_2\}
  \end{align*}
  and let $b_1$ and $b_2$ be instantiated with predicate variables.
  We leave the formal justification of such refinement-polymorphic types as future work.
\end{remark}

\section{Complete Definitions and Proofs}
\label{sec:fulldef}
\subsection{Well-Formedness of Types}

\begin{figure}[h]
  \begin{flushleft}
    \fbox{$\Delta\wf\tau$}
  \end{flushleft}
  \begin{minipage}{0.47\hsize}
    \infrule{
      \Delta,x:B\wf\pred:\bool
    }{
      \Delta\wf \{x:B\mid\pred\}
    }
  \end{minipage}
  \begin{minipage}{0.47\hsize}
    \infrule{
      \Delta,\ST{x:\tau_1}\wf\tau_2
    }{
      \Delta\wf x{:}\tau_1\to\tau_2
    }
  \end{minipage}

  \begin{flushleft}
    \fbox{$\Gamma\st t:B$}
  \end{flushleft}
  \begin{minipage}{0.2\hsize}
    \infax{
      \Delta\st x:\Delta(x)
    }
  \end{minipage}
  \begin{minipage}{0.23\hsize}
    \infax{
      \Delta\st\code{true}:\bool
    }
  \end{minipage}
  \begin{minipage}{0.23\hsize}
    \infax{
      \Delta\st\code{false}:\bool
    }
  \end{minipage}
  \begin{minipage}{0.2\hsize}
    \infax{
      \Delta\st n:\int
    }
  \end{minipage}

  \vspace{5pt}
  \begin{minipage}{0.3\hsize}
    \infrule{
      \Delta\st s_i:\int\quad(i=1,2)
    }{
      \Delta\st s_1=s_2:\bool
    }
  \end{minipage}
  \begin{minipage}{0.4\hsize}
    \infrule{
      \Delta\st S_i:\int\li \quad (i=1,2)
    }{
      \Delta\st S_1=S_2:\bool
    }
  \end{minipage}
  \begin{minipage}{0.2\hsize}
    \infrule{
      \Delta\st\pred:\bool
    }{
      \Delta\st\lnot\pred:\bool
    }
  \end{minipage}

  \vspace{5pt}
  \begin{minipage}{0.3\hsize}
    \infrule{
      \Delta\st\pred_i:\bool \quad (i=1,2)
    }{
      \Delta\st\lnot\pred_1\land\pred_2:\bool
    }
  \end{minipage}
  \begin{minipage}{0.3\hsize}
    \infrule{
      \Delta\st\pred_i:\bool \quad (i=1,2)
    }{
      \Delta\st\lnot\pred_1\lor\pred_2:\bool
    }
  \end{minipage}

  \vspace{5pt}
  \begin{minipage}{0.4\hsize}
    \infrule{
      \Delta\st S_i:\int\li\quad (i=1,2)
    }{
      \Delta\st\code{broadcastable}(S_1,S_2):\bool
    }
  \end{minipage}
  \begin{minipage}{0.4\hsize}
    \infrule{
      \Delta\st S_i:\int\li\quad (i=1,2)
    }{
      \Delta\st\code{reshapeable}(S_1,S_2):\bool
    }
  \end{minipage}

  \vspace{10pt}
  \begin{minipage}{0.29\hsize}
    \infrule{
      \Delta\st s_i:\int (i=1,\ldots,n)
    }{
      \Delta\st [s_1;\ldots;s_n]:\int\li
    }
  \end{minipage}
  \begin{minipage}{0.4\hsize}
    \infrule{
      \Delta\st S_i:\int\li\quad (i=1,2)
    }{
      \Delta\st\code{append}(S_1,S_2):\int\li
    }
  \end{minipage}
  \begin{minipage}{0.29\hsize}
    \infrule{
      \Delta(x) = \tensor
    }{
      \Delta\st \shape{x}:\int\li
    }
  \end{minipage}

  \vspace{5pt}
  \begin{minipage}{0.3\hsize}
    \infrule{
      \Delta\st S:\int\li
    }{
      \Delta\st\len{S}:\int
    }
  \end{minipage}
  \begin{minipage}{0.4\hsize}
    \infrule{
      \Delta\st s:\int
      \andalso \Delta\st S:\int\li
    }{
      \Delta\st \nth{s,S} : \int
    }
  \end{minipage}
  \begin{minipage}{0.2\hsize}
    \infrule{
      \Delta\st s:\int
    }{
      \Delta\st -s:\int
    }
  \end{minipage}

  \vspace{5pt}
  \begin{minipage}{0.3\hsize}
    \infrule{
      \Delta\st s_i:\int\quad (i=1,2)
    }{
      \Delta\st s_1+s_2:\int
    }
  \end{minipage}
  \begin{minipage}{0.3\hsize}
    \infrule{
      \Delta\st s_i:\int\quad (i=1,2)
    }{
      \Delta\st s_1\times s_2:\int
    }
  \end{minipage}
  \begin{minipage}{0.3\hsize}
    \infrule{
      \Delta\st s_i:\int\quad (i=1,2)
    }{
      \Delta\st \frac{s_1}{s_2}:\int
    }
  \end{minipage}
  \caption{Complete rules for the typing rules of predicates, shapes and sizes.}
  \label{fig:well-formed-}
\end{figure}

\subsection{Semantics of Cast Terms}

\begin{figure}[h]
  \fbox{$[v/x]N$}
  \begin{alignat*}{2}
    [v/x]c &= c \\
    [v/x]y &= \begin{cases}
      v & (x = y) \\
      y & (x \neq y) \\
    \end{cases} \\
    [v/x](N\,v') &= ([v/x]N)\,([v/x]v') \\
    [v/x](\lambda y^\tau.N) &= \lambda y^{[v/x]\tau}. [v/x] N \\
    [v/x](\fix{f^{\tau},y,N}) &= \fix{f^{[v/x]\tau},y,[v/x]N} \\
    [v/x](\assert{\pred}; N) &= \assert{[v/x]\pred}; [v/x]N \\
    [v/x](\ifelse{v_1}{N_1}{N_2}) &= \ifelse{[v/x]v_1}{[v/x]N_1}{[v/x]N_2} \\
    [v/x](\letin{y^{\tau}=N_1}N_2) &= (\letin{y^{[v/x]\tau}=[v/x]N_1}[v/x]N_2)
  \end{alignat*}
  (We assume variables are appropriately alpha-renamed so that variables at different scopes do not collide)

  \vspace{10pt}
  \fbox{$N_1\reduce N_2$}
  \begin{alignat*}{2}
    \assert{\code{true}};N &\reduce N \\
    \assert{\code{false}};N &\reduce \blame \\
    (\lambda x^{\tau}.N_1)\,v &\reduce [v/x]N_1 \\
    (\fix{f^\tau,x,N_1})\,v &\reduce [v/x,\fix{f^\tau,x,N_1}/f]N_1 \\
    c\,v &\reduce \code{ev}(c,v) \\
    \letin{x^{\tau}=v} N &\reduce [v/x]N \\
    \ifelse{\code{true}}{N_1}{N_2} &\reduce N_1 \\
    \ifelse{\code{false}}{N_1}{N_2} &\reduce N_2 \\
    C[N_1] &\reduce \begin{cases}
      C[N_2] & (N_1 \reduce N_2) \\
      \blame & (N_1 \reduce \blame) \\
    \end{cases}
  \end{alignat*}
  \begin{align*}
    C \, \text{(context)}
    &::= \square \mid C\,N \mid v\,C \mid \letin{x=C}N
  \end{align*}

  \caption{Substitution and reduction of the target language (full version of Figure~\ref{fig:reduction}).}
  \label{fig:reduction-}
\end{figure}

Figure~\ref{fig:reduction-} defines the full definition of the reduction of the cast terms $N_1\reduce N_2$.

\subsection{Properties about Type System}

\begin{proposition}\label{prop:subtyping-lesssim-}(Proposition~\ref{prop:subtyping-lesssim} in paper)
  $\Gamma;\pred\p \tau_1<:\tau_2$ implies $\Gamma;\pred\p\tau_1\lesssim\tau_2\leadsto N$ for some $N$ where all the assertions in $N$ are of the form $\assert{\code{true}}; N'$.
\end{proposition}

\begin{proof}
  By induction on $\Gamma;\pred\p\tau_1<:\tau_2$.
  \begin{itemize}
    \item Case \rn{(Sub-Base)}.
    \infrule{
      \vDash\forall\ST{\Gamma},x{:}B.\refine{\Gamma}\land\pred\land\pred_1\Rightarrow\pred_2
    }{
      \Gamma;\pred\p \{x:B\mid\pred_1\} <: \{x:B\mid\pred_2\}
    }
    Since $\vDash\exists\ST{\Gamma},x{:}B.\refine{\Gamma}\land\pred\land\pred_1\land\pred_2$ holds,
    the following is obtained as expected.
    \[
      \Gamma;\pred\p\{x:B\mid\pred_1\}\lesssim\{x:B\mid\pred_2\}\leadsto\lambda x^{\{x:B\mid\pred_1\}}.\assert{\code{true}};x
    \]
    \item Case \rn{(Sub-Fun)}.
    \infrule{
      \Gamma;\pred\p \tau_5 <: \tau_3
      \andalso \Gamma,x:\tau_5;\pred\p\tau_4 <: \tau_6
    }{
      \Gamma;\pred\p x{:}\tau_3\to\tau_4 <: x{:}\tau_5\to\tau_6
    }
    From the induction hypothesis, there exists $N_1$ and $N_2$ such that $\Gamma;\pred\p\tau_5\lesssim\tau_3\leadsto N_1$ and $\Gamma,x:\tau_5;\pred\p\tau_4\lesssim\tau_6\leadsto N_2$ hold, and all the assertions in $N_1$ and $N_2$ are of the form $\assert{\code{true}};N'$.
    Since $\Gamma,x:\tau_3\sqcap\tau_5\sqsubseteq\Gamma,x:\tau_5$ holds,
    $\Gamma,x:\tau_3\sqcap\tau_5;\pred\p\tau_4\lesssim\tau_6\leadsto N_2$ follows.
    Therefore, $\Gamma;\pred\p x{:}\tau_3\to\tau_4\lesssim x{:}\tau_5\to\tau_6 \leadsto N$ holds for an $N$ such that $N\equiv\lambda f^{x{:}\tau_3\to\tau_4}.\lambda x^{\tau_5}.\letin{x^{\tau_3\sqcap\tau_5}=N_1\,x}\letin{y^{\tau_4}=f\,x}N_2\,y$.
  \end{itemize}
\end{proof}

\subsection{Type Safety}

\begin{lemma}\label{lemma:cast-term-type-prep}
  Let $\BTE{\Gamma}$ and $\self{\tau,x}$ be defined as follows.
  \begin{align*}
    \BTE{\emptyset} &= \emptyset \\
    \BTE{\Gamma,x:\{y:B\mid\pred\}} &= \BTE{\Gamma},x:\{y:B\mid\pred\} \\
    \BTE{\Gamma,x:y{:}\tau_1\to\tau_2} &= \BTE{\Gamma} \\
    \self{\{x:B\mid\pred\}, y} &= \{x:B\mid\pred\land x=y\} \\
    \self{x{:}\tau_1\to\tau_2, y} &= x{:}\tau_1\to\tau_2
  \end{align*}
  Then, $\Gamma;\pred\p\tau_1\lesssim\tau_2\leadsto N$ implies
  $\BTE{\Gamma};\pred\p N:x{:}\tau_1\to\self{\tau_2,x}$ for a variable $x$ does not occur in $\tau_2$.
\end{lemma}

\begin{proof}
  By induction on the derivation of $\Gamma;\pred\p\tau_1\lesssim\tau_2\leadsto N$.
  \begin{itemize}
    \item Case \rn{(Cast-Base)}.
    The derivation must be of the following form (note $\tau_1\equiv\{x:B\mid\pred_1\}$ and $\tau_2\equiv\{x:B\mid\pred_2\}$).
    \infrule{
      \vDash\exists\ST{\Gamma},x{:}B. \refine{\Gamma}\land\pred\land\pred_1\land\pred_2 \\
      \vDash\forall\ST{\Gamma},x{:}B. \refine{\Gamma}\land\pred\land\pred_1\Rightarrow(\pred'\Iff\pred_2)
    }{
      \Gamma;\pred\p \{x:B\mid\pred_1\} \lesssim \{x:B\mid\pred_2\}
      \leadsto \lambda x^{\{x:B\mid\pred_1\}}.\assert{\pred'}; x
    }
    Noting $\ST{\Gamma}=\ST{\BTE{\Gamma}}$ and $\refine{\Gamma}=\refine{\BTE{\Gamma}}$, the following holds.
    \[
      \vDash\forall\ST{\BTE{\Gamma}},x{:}B.\refine{\BTE{\Gamma}}\land\pred\land\pred_1\land\pred'\land y=x\Rightarrow[y/x]\pred_2
    \]
    Therefore, the following proof tree concludes the proof.
    \begin{prooftree}
      \AxiomC{
        \stackanchor
          {$\BTE{\Gamma},x:\tau_1;\pred\land\pred'\p x:\{y:B\mid y=x\}$}
          {$\BTE{\Gamma},x:\tau_1;\pred\land\pred'\p \{y:B\mid y=x\}<:\{y:B\mid[y/x]\pred_2\land y=x\}$}
      }
      \UnaryInfC{$\BTE{\Gamma},x:\tau_1;\pred\p\assert{\pred'}; x:\{y:B\mid[y/x]\pred_2\land y=x\}$}
      \UnaryInfC{$\BTE{\Gamma};\pred\p(\lambda x^{\{x:B\mid\pred_1\}}.\assert{\pred'}; x):x{:}\tau_1\to\{y:B\mid[y/x]\pred_2\land y=x\}$}
    \end{prooftree}

    \item Case \rn{(Cast-Fun)}.
    The derivation must be of the following form (note $\tau_1\equiv x{:}\tau_3\to\tau_4$ and $\tau_2\equiv x{:}\tau_5\to\tau_6$).
    \infrule{
      \Gamma;\pred\p \tau_5\lesssim\tau_3 \leadsto N_1 \andalso
      \Gamma,x:\tau_3\sqcap\tau_5;\pred\p\tau_4\lesssim\tau_6 \leadsto N_2
    }{
      \Gamma;\pred\p x{:}\tau_3\to\tau_4 \lesssim x{:}\tau_5\to\tau_6 \\
      \leadsto \lambda f^{x{:}\tau_3\to\tau_4}.\lambda x^{\tau_5}. (\letin{y^{\tau_3\sqcap\tau_5}=N_1\,x}\letin{z^{\tau_4}=f\,y}N_2\,z)
    }
    The following holds from the induction hypothesis for some variables $a$ and $b$ that does not occur in $\tau_3$ and $\tau_6$ respectively.
    \begin{gather}
     \BTE{\Gamma};\pred\p N_1:a{:}\tau_5\to\self{\tau_3,a} \label{eq:cast-term-type-prep-eq1} \\
     \BTE{\Gamma,x:\tau_3\sqcap\tau_5};\pred\p N_2:b{:}\tau_4\to\self{\tau_6,b} \label{eq:cast-term-type-prep-eq2}
    \end{gather}
    The following proof tree concludes the proof ($\Gamma':=\BTE{\Gamma},f:(x{:}\tau_3\to\tau_4),x:\tau_5$).
    \begin{prooftree}
      \AxiomC{$\Pi_1$}
      \UnaryInfC{$\Gamma';\pred\p N_1\,x:\self{\tau_3,x}$}
      \AxiomC{$\Pi_2$}
      \UnaryInfC{$\Gamma',y{:}\self{\tau_3,x};\pred\p (\letin{z^{\tau_4}=f\,y}N_2\,z):\tau_6$}
      \BinaryInfC{$\Gamma';\pred\p(\letin{y^{\tau_3\sqcap\tau_5}=N_1\,x}\letin{z^{\tau_4}=f\,y}N_2\,z):x{:}\tau_5\to\tau_6$}
      \UnaryInfC{
        \stackanchor{$\BTE{\Gamma};\pred\p \lambda f^{x{:}\tau_3\to\tau_4}.\lambda x^{\tau_5}. (\letin{y^{\tau_3\sqcap\tau_5}=N_1\,x}\letin{z^{\tau_4}=f\,y}$}{$N_2\,z):f{:}(x{:}\tau_3\to\tau_4)\to(x{:}\tau_5\to\tau_6)$}}
    \end{prooftree}
    $\Pi_1$:
    \begin{prooftree}
      \AxiomC{$\Gamma';\pred\p N_1: a{:}\tau_5\to\self{\tau_3,a}$}
      \AxiomC{$\Gamma';\pred\p x:\tau_5$}
      \BinaryInfC{$\Gamma';\pred\p N_1\,x:\self{\tau_3,x}$}
    \end{prooftree}
    $\Pi_2$:
    \begin{prooftree}
      \AxiomC{$\Pi_3$}
      \UnaryInfC{$\Gamma',y{:}\self{\tau_3,x};\pred\p f\,y:\tau_4$}
      \AxiomC{
        \stackanchor
          {$\Gamma',y{:}\self{\tau_3,x},z{:}\tau_4;\pred\p N_2: b{:}\tau_4\to\self{\tau_6,b}$}
          {$\Gamma',y{:}\self{\tau_3,x},z{:}\tau_4;\pred\p z:\tau_4$}
      }
      \UnaryInfC{$\Gamma',y{:}\self{\tau_3,x},z{:}\tau_4;\pred\p N_2\,z:\self{\tau_6,z}$}
      \UnaryInfC{$\Gamma',y{:}\self{\tau_3,x},z{:}\tau_4;\pred\p N_2\,z:\tau_6$}
      \BinaryInfC{$\Gamma',y{:}\self{\tau_3,x};\pred\p (\letin{z^{\tau_4}=f\,y}N_2\,z):\tau_6$}
    \end{prooftree}
    $\Pi_3$:
    \begin{prooftree}
      \AxiomC{
        \stackanchor
          {$\Gamma',y{:}\self{\tau_3,x};\pred\p f: x{:}\tau_3\to\tau_4$}
          {$\Gamma',y{:}\self{\tau_3,x};\pred\p y:\tau_3$}
      }
      \UnaryInfC{$\Gamma',y{:}\self{\tau_3,x};\pred\p f\,y:[y/x]\tau_4$}
      \AxiomC{$\Gamma',y{:}\self{\tau_3,x};\pred\p [y/x]\tau_4<:\tau_4$}
      \BinaryInfC{$\Gamma',y{:}\self{\tau_3,x};\pred\p f\,y:\tau_4$}
    \end{prooftree}

    In $\Pi_3$, one of the leaves $\Gamma',y{:}\self{\tau_3,x};\pred\p [y/x]\tau_4<:\tau_4$ is proven as follows:
    \begin{itemize}
      \item If $\tau_3$ is a function type, so is $\tau_5$, and $x$ is a variable of function type.
      Therefore, $x$ does not occur in $\tau_4$, and $[y/x]\tau_4=\tau_4$ holds.
      \item If $\tau_3$ is a base type, $\refine{\Gamma',y{:}\self{\tau_3,x}}$ implies $y = x$. Therefore, \\ $\Gamma';y{:}\self{\tau_3,x};\pred\p [y/x]\tau_4<:\tau_4$ holds.
    \end{itemize}

    \vspace{5pt}
    Also, in $\Pi_2$, one of the leaves
    \begin{equation}
      \Gamma',y{:}\self{\tau_3,x},z{:}\tau_4;\pred\p N_2:b{:}\tau_4\to\self{\tau_6,b} \label{eq:cast-term-type-prep-eq3}
    \end{equation}
    is proven as follows.
    \begin{itemize}
      \item If $\tau_3$ is a function type, so is $\tau_3\sqcap\tau_5$, and \eqref{eq:cast-term-type-prep-eq2} is equivalent to $\BTE{\Gamma};\pred\p N_2:b{:}\tau_4\to\self{\tau_6,b}$.
      Therefore, \eqref{eq:cast-term-type-prep-eq3} holds trivially.
      \item If $\tau_3$ is a base type, so is $\tau_5$.
      Let $\tau_3\equiv\{x:B\mid\pred_3\}$ and $\tau_5\equiv\{x:B\mid\pred_5\}$.
      Then, $\refine{\Gamma',y{:}\self{\tau_3,x}}$ is logically equivalent to $\refine{\BTE{\Gamma,x{:}\tau_3\sqcap\tau_5}}$ as follows:
      \begin{align*}
        \refine{\Gamma',y{:}\self{\tau_3,x}}
        &= \refine{\BTE{\Gamma}}\land\pred_5\land[y/x]\pred_3\land y=x \\
        \refine{\BTE{\Gamma,x{:}\tau_3\sqcap\tau_5}}
        &= \refine{\BTE{\Gamma}}\land\pred_3\land\pred_5
      \end{align*}
      Therefore, $\Gamma',y{:}\self{\tau_3,x};\pred\p N_2:b{:}\tau_4\to\self{\tau_6,b}$ holds, and so does \eqref{eq:cast-term-type-prep-eq3}.
    \end{itemize}
  \end{itemize}
\end{proof}

\begin{lemma}\label{lemma:translation-preserves-type-}(Lemma~\ref{lemma:translation-preserves-type} in paper)
  $\Gamma;\pred\p M\leadsto N:\tau$ implies $\Gamma;\pred\p N:\tau$.
\end{lemma}

\begin{proof}
  By induction on the derivation of $\Gamma;\pred\p M\leadsto N:\tau$.
  \begin{itemize}
    \item Case \rn{(CI-App)}.
    \infrule{
      \Gamma;\pred\p M_1\leadsto N_1:y{:}\tau_1\to\tau_2
      \andalso \Gamma(x)=\tau_3
      \andalso \Gamma;\pred\p \tau_3 \lesssim \tau_1 \leadsto N_2
    }{
      \Gamma;\pred\p M_1\,x \leadsto (\letin{x^{\tau_1}=N_2\,x}N_1\,x):[x/y]\tau_2
    }
    From the induction hypothesis and Lemma~\ref{lemma:cast-term-type}, we obtain the following for a fresh $z$.
    \begin{gather*}
      \Gamma;\pred\p N_1:y{:}\tau_1\to\tau_2
      \andalso \Gamma;\pred\p N_2:z{:}\tau_3\to\tau_1
    \end{gather*}
    Noting $[x/z]\tau_1=\tau_1$, it follows that $\Gamma;\pred\p N_2\,x:\tau_1$ holds, and $\Gamma;\pred\p(\letin{x^{\tau_1}=N_2\,x}N_1\,x):[x/y]\tau_2$ is obtained as expected.

    \item Case \rn{(CI-Sub)}.
    \infrule{
      \Gamma;\pred\p M_1 \leadsto N_1 : \tau_1
      \andalso \Gamma;\pred\p \tau_1 \lesssim \tau \leadsto N_2
    }{
      \Gamma;\pred\p M_1 \leadsto \letin{x^{\tau_1}=N_1}N_2\,x : \tau
    }
    From the induction hypothesis and Lemma~\ref{lemma:cast-term-type}, we obtain $\Gamma;\pred\p N_1:\tau_1$ and $\Gamma;\pred\p N_2:y{:}\tau_1\to\tau$ for a fresh variable $y$.
    Since it is safe to assume that $x$ does not appear in $N_2$ and $y\neq x$, $\Gamma,x:\tau_1;\pred\p N_2:y{:}\tau_1\to\tau$ also holds.
    Thus, $\Gamma,x:\tau_1;\pred\p N_2\,x:\tau$ holds, and hence we obtain the result as follows (note $[x/y]\tau = \tau$) .
    \begin{prooftree}
      \AxiomC{$\Gamma;\pred\p N_1:\tau_1$}
      \AxiomC{$\Gamma,x:\tau_1;\pred\p N_2:y{:}\tau_1\to\tau$}
      \AxiomC{$\Gamma,x:\tau_1;\pred\p x:\tau_1$}
      \BinaryInfC{$\Gamma,x:\tau_1;\pred\p N_2\,x:[x/y]\tau$}
      \BinaryInfC{$\Gamma;\pred\p \letin{x^{\tau_1}=N_1}N_2\,x : \tau$}
    \end{prooftree}

    \item Other cases are trivial.
  \end{itemize}
\end{proof}

\begin{figure}
  \begin{minipage}[t]{0.50\textwidth}
  \fbox{$[v/x]\tau$}
  \begin{align*}
    [v/x]\{y:B\mid\pred\} &= \{y:B\mid[v/x]\pred\} \\
    [v/x](y{:}\tau_1\to\tau_2) &= y{:}([v/x]\tau_1)\to([v/x]\tau_2)
  \end{align*}
  \end{minipage}
  \begin{minipage}[t]{0.48\textwidth}
  \fbox{$[v/x]\Gamma$}
  \begin{align*}
    [v/x]\emptyset &= \emptyset \\
    [v/x](\Gamma,x:\tau) &= [v/x]\Gamma \\
    [v/x](\Gamma,y:\tau) &= ([v/x]\Gamma),y:[v/x]\tau \,\, (x \neq y)
  \end{align*}
  \end{minipage}
  \caption{Substitution of type and type environment.}
  \label{fig:subst-type}
\end{figure}

\begin{definition}
  Substitution of type $[v/x]\tau$ and type environment $[v/x]\Gamma$ is defined in Figure~\ref{fig:subst-type}.
\end{definition}

\begin{lemma}\label{lemma:substitution-preserves-subtyping}
  $\Gamma;\pred\p\tau<:\tau'$ and $\p v:\Gamma(x)$ imply $[v/x]\Gamma;[v/x]\pred\p[v/x]\tau<:[v/x]\tau'$.
\end{lemma}

\begin{proof}
  By induction on $\Gamma;\pred\p\tau<:\tau'$.
  \begin{itemize}
    \item Case \rn{(Sub-Base)}.
    \infrule{
      \vDash\forall\ST{\Gamma},y{:}B.\refine{\Gamma}\land\pred\land\pred_1\Rightarrow\pred_2
    }{
      \Gamma;\pred\p \{y:B\mid\pred_1\} <: \{y:B\mid\pred_2\}
    }
    We can assume $x\neq y$ w.l.o.g by alpha renaming.
    If $\Gamma(x)$ is a function type, $x\not\in\dom{\ST{\Gamma}}$ and $x$ does not appear in $\refine{\Gamma}$, $\pred$, $\pred_1$ or $\pred_2$.
    Therefore, $\ST{[v/x]\Gamma}=\ST{\Gamma}$ and $\refine{[v/x]\Gamma}=\refine{\Gamma}$, we obtain $\forall\ST{[v/x]\Gamma},y{:}B.\refine{[v/x]\Gamma}\land[v/x]\pred\land[v/x]\pred_1\Rightarrow[v/x]\pred_2$.
    Therefore, $[v/x]\Gamma;[v/x]\pred\p [v/x]\{y:B\mid\pred_1\} <: [v/x]\{y:B\mid\pred_2\}$ holds.

    If $\Gamma(x)$ is a base type, $x\in\dom{\ST{\Gamma}}$ and
    $\dom{\ST{[v/x]\Gamma}}=\dom{\ST{\Gamma}}\setminus\{x\}$.
    Since $v$ is a closed term,
    we obtain $\forall\ST{[v/x]\Gamma},y{:}B.[v/x](\refine{\Gamma}\land\pred\land\pred_1)\Rightarrow[v/x]\pred_2$.
    Therefore, $[v/x]\Gamma;[v/x]\pred\p\{y:B\mid[v/x]\pred_1\}<:\{y:B\mid[v/x]\pred_2\}$ holds.

    \item Case \rn{(Sub-Fun)}.
    \infrule{
      \Gamma;\pred\p \tau_3 <: \tau_1
      \andalso \Gamma,y:\tau_3;\pred\p\tau_2 <: \tau_4
    }{
      \Gamma;\pred\p y{:}\tau_1\to\tau_2 <: y{:}\tau_3\to\tau_4
    }
    We can assume $x\neq y$ w.l.o.g by alpha renaming.
    From the induction hypothesis, we obtain:
    \begin{gather*}
      [v/x]\Gamma;[v/x]\pred\p[v/x]\pred_3<:[v/x]\pred_1 \\
      ([v/x]\Gamma),x:[v/x]\tau_3;[v/x]\pred\p[v/x]\tau_2<:[v/x]\tau_4
    \end{gather*}
    Therefore, $[v/x]\Gamma;[v/x]\pred\p [v/x](y{:}\tau_1\to\tau_2)<:[v/x](y{:}\tau_3\to\tau_4)$ holds as expected.
  \end{itemize}
\end{proof}

\begin{lemma}\label{lemma:substitution-preserves-typing}
  $\Gamma;\pred\p N:\tau$ and $\p v:\Gamma(x)$ imply $[v/x]\Gamma;[v/x]\pred\p [v/x]N:[v/x]\tau$.
\end{lemma}

\begin{proof}
  By induction on $\Gamma;\pred\p N:\tau$.
  \begin{itemize}
    \item Case \rn{(CT-Con)}, \rn{(CT-VF)} and \rn{(CT-VB)} is trivial.
    \item Case \rn{(CT-Lam)}.
    \infrule{
      \Gamma,y:\tau_1;\pred\p N_1:\tau_2
    }{
      \Gamma;\pred\p \lambda y^{\tau_1}.N_1 : y{:}\tau_1\to\tau_2
    }
    We can assume $x \neq y$ by alpha renaming.
    From the induction hypothesis,
    \[
      ([v/x]\Gamma),y:[v/x]\tau_1;[v/x]\pred\p[v/x]N_1:[v/x]\tau_2
    \]
    holds.
    Therefore, $[v/x]\Gamma;[v/x]\pred\p\lambda y^{[v/x]\tau_1}.[v/x]N_1:y{:}([v/x]\tau_1)\to[v/x]\tau_2$ holds as expected.

    \item Case \rn{(CT-Fix)} is similar to the case \rn{(CT-Lam)}.

    \item Case \rn{(CT-App)}.
    \infrule{
      \Gamma;\pred\p N_1:y{:}\tau_1\to\tau_2
      \andalso \Gamma;\pred\p v_1:\tau_1
    }{
      \Gamma;\pred\p N_1\,v_1:[v_1/y]\tau_2
    }
    We assume $x \neq y$ by alpha renaming.
    From the induction hypothesis, we obtain the following.
    \begin{gather*}
      [v/x]\Gamma;[v/x]\pred\p[v/x]N_1:y{:}[v/x]\tau_1\to[v/x]\tau_2 \\
      [v/x]\Gamma;[v/x]\pred\p[v/x]v_1:[v/x]\tau_1
    \end{gather*}
    Therefore, $[v/x]\Gamma;[v/x]\pred\p ([v/x]N_1)\,([v/x]v_1):[[v/x]v_1/y][v/x]\tau_2$ holds as expected.

    \item Case \rn{(CT-Sub)}.
    \infrule{
      \Gamma;\pred\p N:\tau'
      \andalso \Gamma;\pred\p \tau'<:\tau
    }{
      \Gamma;\pred\p N:\tau
    }
    From the induction hypothesis and Lemma~\ref{lemma:substitution-preserves-subtyping},
    we obtain the following.
    \begin{gather*}
      [v/x]\Gamma;[v/x]\pred\p[v/x]N:[v/x]\tau' \\
      [v/x]\Gamma;[v/x]\pred\p[v/x]\tau'<:[v/x]\tau
    \end{gather*}
    Therefore,
    $[v/x]\Gamma;[v/x]\pred\p[v/x]N:[v/x]\tau$ follows as expected.

    \item Other cases are trivial from the induction hypothesis.
  \end{itemize}
\end{proof}

From now on, we abbreviate $\emptyset;\top\p N:\tau$ as $\p N:\tau$.

\begin{lemma}[Progress]\label{lemma:type-progress}
  Suppose $\p N:\tau$ holds, and $\evof{c,v}$ is defined whenever $\p c\,v:\tau$ holds for some $\tau$.
  Then, one of the following holds.
  \begin{itemize}
    \item $N$ is a value.
    \item $N \reduce N'$ for some $N'$
    \item $N \reduce \blame$
  \end{itemize}
\end{lemma}

\begin{proof}
  By induction on the derivation of $\p N:\tau$.
  \begin{itemize}
    \item In cases \rn{(CT-Con)}, \rn{(CT-VF)}, \rn{(CT-VB)}, \rn{(CT-Lam)} and \rn{(CT-Fix)}, $N$ is a value.
    \item Case \rn{(CT-App)}.
    \infrule{
      \p N_1:x{:}\tau_1\to\tau_2
      \andalso \p v:\tau_1
    }{
      \p N_1\,v:[v/x]\tau_2
    }
    We split cases from the induction hypothesis.
    \begin{itemize}
      \item If $N_1$ is a value, then $N_1$ is either a constant, a lambda term, or a recursive function.
      If $N_1$ is a constant $c$, then $c\,v\reduce\evof{c,v}$ holds since $\evof{c,v}$ is defined whenever $c\,v$ is well-typed.
      If $N_1$ is a lambda term or a recursive function, there exists $N'$ such that $N_1\,v\reduce N'$ holds.
      \item If $N_1\reduce N_1'$ holds for some $N_1'$, then $N_1\,v\reduce N_1'\,v$.
      \item If $N_1\reduce\blame$, then $N_1\,v\reduce\blame$.
    \end{itemize}
    \item Case \rn{(CT-Ass)}.
    \infrule{
      \emptyset;\pred\p N_1:\tau
    }{
      \p \assert{\pred};N_1 : \tau
    }
    Since $\pred$ is a closed predicate, $\pred$ is either $\code{true}$ or $\code{false}$.
    If $\pred\equiv\code{true}$, then $\assert{\pred};N_1\reduce N_1$.
    Otherwise, $\pred\equiv\code{false}$ holds, thus $\assert{\pred};N_1\reduce\blame$.
    \item Case \rn{(CT-If)}.
    \infrule{
      \p v:\{x:\bool\mid\pred\} \andalso
      \emptyset;v\p N_1:\tau
      \andalso \emptyset;\lnot v\p N_2:\tau
    }{
      \p\ifelse{v}{N_1}{N_2}:\tau
    }
    Since $v$ is a closed boolean value, $v$ is either $\top$ or \code{false}.
    If $v\equiv\code{true}$, then $\ifelse{\top}{N_1}{N_2}\reduce N_1$.
    If $v\equiv\code{false}$, then $\ifelse{\code{false}}{N_1}{N_2}\reduce N_2$.
    \item Case \rn{(CT-Let)}.
    \infrule{
      \p N_1:\tau_2
      \andalso x:\tau_2;\top\p N_2:\tau
    }{
      \p \letin{x^{\tau_1}=N_1}N_2 : \tau
    }
    We split cases from the induction hypothesis.
    \begin{itemize}
      \item If $N_1$ is a value, let $v_1$ such that $N_1\equiv v_1$.
      Then, $\letin{x^{\tau_1}=v_1}N_2\reduce[v_1/x]N_2$ holds.
      \item If $N_1\reduce N_1'$ holds for some $N_1'$, then $\letin{x^{\tau_1}=N_1}N_2\reduce\letin{x^{\tau_1}N_1'}N_2$ holds.
      \item If $N_1\reduce\blame$ holds, then $\letin{x^{\tau_1}=N_1}N_2\reduce\blame$.
    \end{itemize}

    \item Case \rn{(CT-Sub)} is immediate from the induction hypothesis.
  \end{itemize}
\end{proof}

\begin{lemma}[Preservation]\label{lemma:type-preservation}
  Suppose $\p c\,v:\tau$ implies $\p \evof{c, v}:\tau$ for every $c, v$ and $\tau$.
  Then, $\p N:\tau$ and $N \reduce N'$ imply $\p N':\tau$.
\end{lemma}

\begin{proof}
  By induction on $N\reduce N'$.
  \begin{itemize}
    \item Case $\assert{\code{true}};N_1\reduce N_1$.
    From the inversion of $\p\assert{\code{true}};N_1:\tau$, it follows that $\p N_1:\tau$.

    \item Case $(\lambda x^{\tau_1}.N_1)\,v\reduce [v/x]N_1$.
    The derivation of $\p(\lambda x^{\tau_1}.N_1)\,v:\tau$ must be of the following form.
    \begin{prooftree}
      \AxiomC{$\p v:\tau_1$}
      \AxiomC{$x:\tau_1;\top\p N_1:\tau_2$}
      \UnaryInfC{$\p (\lambda x^{\tau_1}.N_1):x{:}\tau_1\to\tau_2$}
      \BinaryInfC{$\p(\lambda x^{\tau_1}.N_1)\,v:[v/x]\tau_2$}
    \end{prooftree}
    Using Lemma~\ref{lemma:substitution-preserves-typing}, $x:\tau_1;\top\p [v/x]N_1:[v/x]\tau_2$ holds.
    Since $x$ does not appear freely in $[v/x]N_1$, this entails $\p [v/x]N_1:[v/x]\tau_2$ as expected.
    \item Case $\letin{x^{\tau_1}=v}N_1\reduce [v/x]N_1$ is similar to the previous case.

    \item Case $(\fix{f^{\tau_1},x,N_1})\,v\reduce[v/x,\fix{f^{\tau_1},x,N_1}/v]N_1$.
    The derivation of $\p(\fix{f^{\tau_1},x,N_1})\,v:\tau$ is as follows (note $\tau_1\equiv x{:}\tau_2\to\tau_3$).
    \begin{prooftree}
      \AxiomC{$\p v:\tau_2$}
      \AxiomC{$f:(x{:}\tau_2\to\tau_3),x:\tau_2;\top\p N_1:\tau_3$}
      \UnaryInfC{$\p\fix{f^{x{:}\tau_2\to\tau_3},x,N_1}:x{:}\tau_2\to\tau_3$}
      \BinaryInfC{$\p(\fix{f^{x{:}\tau_2\to\tau_3},x,N_1})\,v:[v/x]\tau_3$}
    \end{prooftree}
    Using Lemma~\ref{lemma:substitution-preserves-typing}, $f:(x{:}\tau_2\to\tau_3),x:\tau_2;\top\p[v/x,\fix{f^{\tau_1},x,N_1}/v]N_1:\tau_3$ holds.
    Therefore,\\ $\p[v/x,\fix{f^{\tau_1},x,N_1}/v]N_1:\tau_3$ holds as expected.

    \item Case $c\,v\reduce\evof{c,v}$ follows from the assumption.
    \item Case $\ifelse{\code{true}}{N_1}{N_2}\reduce N_1$ is trivial.
    \item Case $\ifelse{\code{false}}{N_1}{N_2}\reduce N_2$ is trivial.
    \item Other cases are trivial.
  \end{itemize}
\end{proof}

\begin{theorem}[Type Safety]
  Suppose $\p c\,v:\tau$ implies $\p \evof{c, v}:\tau$ for every $c, v$ and $\tau$.
  Then, $\emptyset;\code{true}\p M \leadsto N : \tau$ implies either $N \reduce^* v, N \Uparrow$ or $N \reduce^* \blame$.
\end{theorem}

\begin{proof}
  From Lemma~\ref{lemma:translation-preserves-type-}, $\p N:\tau$ holds.
  Therefore, the result is obtained from Lemma~\ref{lemma:type-progress} and Lemma~\ref{lemma:type-preservation}.
\end{proof}

\subsection{Properties about Precision}

\begin{lemma}\label{lemma:meet-precision}
  $\seq{x}\p\tau_1\sqsubseteq\tau_2$ and $\seq{x}\p\tau_3\sqsubseteq\tau_4$
  imply $\seq{x}\p\tau_1\sqcap\tau_3\sqsubseteq\tau_2\sqcap\tau_4$.
\end{lemma}

\begin{lemma}\label{lemma:lesssim-relax-}(Lemma~\ref{lemma:lesssim-relax} in paper)
  $\Gamma;\pred\p\tau_1\lesssim\tau_2$,
  $\dom{\Gamma}\p\tau_1\sqsubseteq\tau_3$,
  $\dom{\Gamma}\p\tau_2\sqsubseteq\tau_4$ and
  $\Gamma\sqsubseteq\Gamma'$ imply
  $\Gamma';\pred\p\tau_3\lesssim\tau_4$.
\end{lemma}

\begin{proof}
  By induction on $\Gamma;\pred\p\tau_1\lesssim\tau_2$.
  \begin{itemize}
    \item Case \rn{(Cast-Base)}. The derivation must be of the following form.
    \infrule{
      \vDash\exists\ST{\Gamma},x{:}B. \refine{\Gamma}\land\pred\land\pred_1\land\pred_2 \\
      \vDash\forall\ST{\Gamma},x{:}B. \refine{\Gamma}\land\pred\land\pred_1\Rightarrow(\pred'\Iff\pred_2)
    }{
      \Gamma;\pred\p \{x:B\mid\pred_1\} \lesssim \{x:B\mid\pred_2\} \leadsto \lambda x^{\{x:B\mid\pred_1\}}.\assert{\pred'};x
    }
    Let $\pred_3$ such that $\tau_3\equiv\{x:B\mid\pred_3\}$, $\pred_4$ such that $\tau_4\equiv\{x:B\mid\pred_4\}$, and $\Gamma'$ such that $\Gamma\sqsubseteq\Gamma'$.
    From $\dom{\Gamma}\p\tau_1\sqsubseteq\tau_3$, $\dom{\Gamma}\p\tau_2\sqsubseteq\tau_4$ and $\Gamma\sqsubseteq\Gamma'$, we obtain the following, noting that only a variable of base type appears in the predicates.
    \begin{gather*}
      \forall\ST{\Gamma},x{:}B.\pred_1\Rightarrow\pred_3 \\
      \forall\ST{\Gamma},x{:}B.\pred_2\Rightarrow\pred_4 \\
      \forall\ST{\Gamma}.\refine{\Gamma}\Rightarrow\refine{\Gamma'}
    \end{gather*}
    Therefore, $\exists\ST{\Gamma'},x{:}B.\refine{\Gamma'}\land\pred\land\pred_3\land\pred_4$ holds, noting $\ST{\Gamma}=\ST{\Gamma'}$.
    Let $\pred''$ such that $\forall\ST{\Gamma'},x{:}B.\refine{\Gamma'}\land\pred\land\pred_3\Rightarrow(\pred''\Iff\pred_4)$.
    Thus, $\Gamma';\pred\p\{x:B\mid\pred_3\}\lesssim\{x:B\mid\pred_4\}\leadsto \lambda x^{\{x:B\mid\pred_3\}}.\assert{\pred''};x$ holds as expected.

    \item Case \rn{(Cast-Fun)}. The derivation must be of the following form.
    \infrule{
      \Gamma;\pred\p \tau_2'\lesssim\tau_1' \leadsto N_1 \andalso
      \Gamma,x:\tau_1'\sqcap\tau_2';\pred\p\tau_1''\lesssim\tau_2'' \leadsto N_2
    }{
      \Gamma;\pred\p x{:}\tau_1'\to\tau_1'' \lesssim x{:}\tau_2'\to\tau_2'' \\
      \leadsto \lambda f^{x{:}\tau_1'\to\tau_1''}.\lambda x^{\tau_2'}.(\letin{x^{\tau_1'\sqcap\tau_2'}=N_1\,x}\letin{y^{\tau_1''}=f\,x}N_2\,y)
    }
    Let $\tau_3',\tau_3'', \tau_4'$ and $\tau_4''$ such that $\tau_3\equiv x{:}\tau_3'\to\tau_3''$ and $\tau_4\equiv x{:}\tau_4'\to\tau_4''$.
    From $\dom{\Gamma}\p\tau_1\sqsubseteq\tau_3$ and $\dom{\Gamma}\p\tau_2\sqsubseteq\tau_4$, the following holds.
    \begin{gather*}
      \dom{\Gamma}\p\tau_1'\sqsubseteq\tau_3'
      \andalso \dom{\Gamma},x\p\tau_1''\sqsubseteq\tau_3'' \\
      \dom{\Gamma}\p\tau_2'\sqsubseteq\tau_4'
      \andalso \dom{\Gamma},x\p\tau_2''\sqsubseteq\tau_4''
    \end{gather*}
    Using Lemma~\ref{lemma:meet-precision}, $\dom{\Gamma}\p\tau_1'\sqcap\tau_2'\sqsubseteq\tau_3'\sqcap\tau_4'$ holds.
    Let $\Gamma'$ such that $\Gamma\sqsubseteq\Gamma'$.
    Then, $\Gamma,x:\tau_1'\sqcap\tau_2'\sqsubseteq\Gamma',x:\tau_3'\sqcap\tau_4'$ holds.
    From the induction hypothesis, $\Gamma';\pred\p\tau_4'\lesssim\tau_3'\leadsto N_3$ and $\Gamma',x:\tau_3'\sqcap\tau_4';\pred\p\tau_3''\lesssim\tau_4''\leadsto N_4$ hold for some $N_3$ and $N_4$.
    Therefore, we obtain $\Gamma;\pred\p x{:}\tau_3'\to\tau_3''\lesssim x{:}\tau_4'\to\tau_4'' \leadsto \lambda f^{x{:}\tau_3'\to\tau_3''}.\lambda x^{\tau_4'}.(\letin{x^{\tau_3'\sqcap\tau_4'}=N_3\,x}\letin{y^{\tau_3''}=f\,x}N_4\,y)$ as expected.
  \end{itemize}
\end{proof}

\begin{lemma}\label{lemma:cast-precision-}(Lemma~\ref{lemma:cast-precision} in paper)
  Suppose $\Gamma\sqsubseteq\Gamma', \dom{\Gamma}\p\tau_1\sqsubseteq\tau_1'$ and $\dom{\Gamma}\p\tau_2\sqsubseteq\tau_2'$.
  Then, $\Gamma;\pred\p\tau_1\lesssim\tau_2\leadsto N$ and $\Gamma';\pred\p\tau_1'\lesssim\tau_2'\leadsto N'$ implies $\Gamma;\pred\p N\sqsubseteq N'$.
\end{lemma}

\begin{proof}
  By induction on the derivation of $\Gamma;\pred\p\tau_1\lesssim\tau_2\leadsto N$.
  \begin{itemize}
    \item Case \rn{(Cast-Base)}.
    \infrule{
      \vDash\exists\ST{\Gamma},x{:}B. \refine{\Gamma}\land\pred\land\pred_1\land\pred_2 \\
      \vDash\forall\ST{\Gamma},x{:}B. \refine{\Gamma}\land\pred\land\pred_1\Rightarrow(\pred'\Iff\pred_2)
    }{
      \Gamma;\pred\p \{x:B\mid\pred_1\} \lesssim \{x:B\mid\pred_2\}
      \leadsto \lambda x^{\{x:B\mid\pred_1\}}.\assert{\pred'};x
    }
    From the inversion of $\dom{\Gamma}\p\tau_1\sqsubseteq\tau_1'$ and $\dom{\Gamma}\p\tau_2\sqsubseteq\tau_2'$, it must be that $\tau_1'\equiv\base{\pred_1'}$ and $\tau_2'\equiv\base{\pred_2'}$ for some $\pred_1'$ and $\pred_2'$ where the following holds.
    \[
      \forall\dom{\Gamma},x.\pred_1\Rightarrow\pred_1' \andalso
      \forall\dom{\Gamma},x.\pred_2\Rightarrow\pred_2'
    \]
    Since the variables of function types cannot be used in the predicates, we can restrict the quantified variables to $\dom{\ST{\Gamma}}$ as follows.
    \begin{gather}
      \forall\dom{\ST{\Gamma}},x.\pred_1\Rightarrow\pred_1' \label{eq:cast-precision-1} \\
      \forall\dom{\ST{\Gamma}},x.\pred_2\Rightarrow\pred_2' \label{eq:cast-precision-2}
    \end{gather}
    The derivation of $\Gamma;\pred\p\tau_1'\lesssim\tau_2'\leadsto N'$ must be of the following form.
    \infrule{
      \vDash\exists \ST{\Gamma},x{:}B. \refine{\Gamma}\land\pred\land\pred_1'\land\pred_2' \\
      \vDash\forall\ST{\Gamma},x{:}B.\refine{\Gamma}\land\pred\land\pred_1'\Rightarrow(\pred''\Iff\pred_2')
    }{
      \Gamma;\pred\p \{x:B\mid\pred_1'\} \lesssim \{x:B\mid\pred_2'\}
      \leadsto \lambda x^{\{x:B\mid\pred_1'\}}.\assert{\pred''};x
    }
    Therefore, we obtain $\forall\ST{\Gamma},x{:}B.\refine{\Gamma}\land\pred\land\pred_1\land\pred'\Rightarrow\pred''$ as follows:
    Assume $\refine{\Gamma}\land\pred\land\pred_1\land\pred'$. Then,  $\pred_1'$ follows from \eqref{eq:cast-precision-1}, and $\pred_2$ follows from the premise of $\Gamma;\pred\p\tau_1\lesssim\tau_2\leadsto N$. Therefore, $\pred_2'$ is obtained by \eqref{eq:cast-precision-2}.
    With the premise of $\Gamma;\pred\p\tau_1'\lesssim\tau_2'\leadsto N'$, we obtain $\pred''$ as expected.

    Thus, we obtain $\Gamma;\pred\p N\sqsubseteq N'$ as follows.
    \begin{prooftree}
      % \AxiomC{$\Gamma\p\base{\pred_1}\sqsubseteq\base{\pred_1'}$}
      \AxiomC{$\forall\ST{\Gamma},x{:}B.\refine{\Gamma}\land\pred\land\pred_1\land\pred'\Rightarrow\pred''$}
      \UnaryInfC{$\Gamma,x:\base{\pred_1};\pred\p\assert{\pred'};x\sqsubseteq\assert{\pred''};x$}
      \UnaryInfC{$\Gamma;\pred\p\lambda x^{\base{\pred_1}}.\assert{\pred'};x\sqsubseteq \lambda x^{\base{\pred_1'}}.\assert{\pred''};x$}
    \end{prooftree}

    \item Case \rn{(Cast-Fun)}.
    \infrule{
      \Gamma;\pred\p \tau_5\lesssim\tau_3 \leadsto N_1
      \andalso \Gamma,x:\tau_3\sqcap\tau_5;\pred\p\tau_4\lesssim\tau_6 \leadsto N_2
    }{
      \Gamma;\pred\p x{:}\tau_3\to\tau_4 \lesssim x{:}\tau_5\to\tau_6 \\
      \leadsto \lambda f^{x{:}\tau_3\to\tau_4}.\lambda x^{\tau_5}. (\letin{x^{\tau_3\sqcap\tau_5}=N_1\,x}\letin{y^{\tau_4}=f\,x}N_2\,y)
    }
    From the inversion of $\dom{\Gamma}\p\tau_1\sqsubseteq\tau_1'$ and $\dom{\Gamma}\p\tau_2\sqsubseteq\tau_2'$, it must be that $\tau_1'\equiv x{:}\tau_3'\to\tau_4'$ and $\tau_2'\equiv x{:}\tau_5'\to\tau_6'$ for some $\tau_3',\tau_4',\tau_5'$ and $\tau_6'$ where the following holds.
    \begin{gather}
      \dom{\Gamma}\p\tau_3\sqsubseteq\tau_3' \label{eq:cast-precision-3} \\
      \dom{\Gamma}\p\tau_5\sqsubseteq\tau_5' \label{eq:cast-precision-4} \\
      \dom{\Gamma},x\p\tau_4\sqsubseteq\tau_4' \label{eq:cast-precision-5} \\
      \dom{\Gamma},x\p\tau_6\sqsubseteq\tau_6' \label{eq:cast-precision-6}
    \end{gather}
    The derivation of $\Gamma';\pred\p\tau_1'\lesssim\tau_2'\leadsto N'$ must be of the following form.
    \infrule{
      \Gamma';\pred\p \tau_5'\lesssim\tau_3' \leadsto N_1'
      \andalso \Gamma',x:\tau_3'\sqcap\tau_5';\pred\p\tau_4'\lesssim\tau_6' \leadsto N_2'
    }{
      \Gamma';\pred\p x{:}\tau_3'\to\tau_4' \lesssim x{:}\tau_5'\to\tau_6' \\
      \leadsto \lambda f^{x{:}\tau_3'\to\tau_4'}.\lambda x^{\tau_5'}. (\letin{x^{\tau_3'\sqcap\tau_5'}=N_1'\,x}\letin{y^{\tau_4'}=f\,x}N_2'\,y)
    }
    With \eqref{eq:cast-precision-3} and \eqref{eq:cast-precision-4}, we obtain $\Gamma;\pred\p N_1\sqsubseteq N_1'$ from the induction hypothesis.
    Noting $\Gamma,x:\tau_3\sqcap\tau_5\sqsubseteq\Gamma',x:\tau_3'\sqcap\tau_5'$, we obtain $\Gamma,x:\tau_3\sqcap\tau_5;\pred\p N_2\sqsubseteq N_2'$ from the induction hypothesis with \eqref{eq:cast-precision-5} and \eqref{eq:cast-precision-6}.
    Thus, $\Gamma;\pred\p N\sqsubseteq N'$ follows as expected.
  \end{itemize}
\end{proof}

\subsection{Gradual Guarantee}

\begin{figure}
  \begin{flushleft}
    \fbox{$\seq{x}\p M_1 \sqsubseteq M_2$}
  \end{flushleft}
  \centering

  \begin{minipage}{0.25\hsize}
    \infax[PM-Const]{
      \seq{x}\p c \sqsubseteq c
    }
  \end{minipage}
  \begin{minipage}{0.25\hsize}
    \infax[PM-Var]{
      \seq{y}\p x \sqsubseteq x
    }
  \end{minipage}
  \begin{minipage}{0.45\hsize}
    \infrule[PM-Lam]{
      \seq{y}\p \tau_1\sqsubseteq\tau_2
      \andalso \seq{y},x\p M_1 \sqsubseteq M_2
    }{
      \seq{y}\p \lambda x{:}\tau_1.M_1 \sqsubseteq \lambda x{:}\tau_2.M_2
    }
  \end{minipage}

  \begin{minipage}{0.55\hsize}
    \infrule[PM-Annot]{
      \seq{y}\p M_1 \sqsubseteq M_2
      \andalso \seq{y}\p \tau_1 \sqsubseteq \tau_2
    }{
      \seq{y}\p (M_1 : \tau_1) \sqsubseteq (M_2 : \tau_2)
    }
  \end{minipage}
  \begin{minipage}{0.4\hsize}
    \infrule[PM-App]{
      \seq{y}\p M_1 \sqsubseteq M_2
    }{
      \seq{y}\p M_1 \, x \sqsubseteq M_2 \, x
    }
  \end{minipage}

  \infrule[PM-Let]{
    \seq{y}\p M_1 \sqsubseteq M_1'
    \andalso \seq{y},x\p M_2 \sqsubseteq M_2'
  }{
    \seq{y}\p (\letin{x = M_1} M_2) \sqsubseteq (\letin{x = M_1'} M_2')
  }

  \infrule[PM-Fix]{
    \seq{y}\p x{:}\tau_1\to\tau_2 \sqsubseteq x{:}\tau_3\to\tau_4
    \andalso \seq{y},f,x\p M \sqsubseteq M'
  }{
    \seq{y}\p \fix{f:(x{:}\tau_1\to\tau_2),x,M} \sqsubseteq \fix{f:(x{:}\tau_3\to\tau_4),x,M'}
  }

  \infrule[PM-If]{
    \seq{y}\p M_1 \sqsubseteq M_1'
    \andalso \seq{y}\p M_2 \sqsubseteq M_2'
  }{
    \seq{y}\p (\ifelse{x}{M_1}{M_2}) \sqsubseteq (\ifelse{x}{M_1'}{M_2'})
  }
  \caption{Precision relation over terms (full version of Figure~\ref{fig:precision-term}).}
  \label{fig:precision-term-}
\end{figure}

\begin{figure}
  \begin{flushleft}
    \fbox{$\Gamma;\pred\p N_1\sqsubseteq N_2$}
  \end{flushleft}

  \begin{minipage}{0.4\hsize}
    \infax[PC-Const]{
      \Gamma;\pred\p c \sqsubseteq c
    }
  \end{minipage}
  \begin{minipage}{0.4\hsize}
    \infax[PC-Var]{
      \Gamma;\pred\p x \sqsubseteq x
    }
  \end{minipage}

  \begin{minipage}{0.5\hsize}
    \infrule[PC-Lam]{
      \dom{\Gamma}\p \tau_1 \sqsubseteq \tau_2
      \andalso \Gamma,x:\tau_1;\pred\p N_1 \sqsubseteq N_2
    }{
      \Gamma;\pred\p\lambda x^{\tau_1}.N_1 \sqsubseteq \lambda x^{\tau_2}.N_2
    }
  \end{minipage}
  \begin{minipage}{0.45\hsize}
    \infrule[PC-App]{
      \Gamma;\pred\p N_1 \sqsubseteq N_2
      \andalso \emptyset;\code{true}\p v_1 \sqsubseteq v_2
    }{
      \Gamma;\pred\p N_1\,v_1 \sqsubseteq N_2\,v_2
    }
  \end{minipage}

  \infrule[PC-Fix]{
    \dom{\Gamma}\p x{:}\tau_1\to\tau_2 \sqsubseteq x{:}\tau_1'\to\tau_2'
    \andalso \Gamma,f:x{:}\tau_1\to\tau_2,x:\tau_1;\pred\p N_1 \sqsubseteq N_2
  }{
    \Gamma;\pred\p \fix{f^{x{:}\tau_1\to\tau_2},x,N_1} \sqsubseteq \fix{f^{x{:}\tau_1'\to\tau_2'},x,N_2}
  }
  \infrule[PC-Let]{
    \Gamma\p N_1\sqsubseteq N_3
    \andalso \Gamma,x:\tau_1;\pred\p N_2 \sqsubseteq N_4
  }{
    \Gamma;\pred\p \letin{x^{\tau_1}=N_1}N_2 \sqsubseteq \letin{x^{\tau_3}=N_3}N_4
  }
  \infrule[PC-If]{
    \emptyset;\code{true}\p v_1 \sqsubseteq v_2
    \andalso \Gamma;\pred\land v_1\p N_1 \sqsubseteq N_3
    \andalso \Gamma;\pred\land\lnot v_1\p N_2 \sqsubseteq N_4
  }{
    \Gamma;\pred\p \ifelse{v_1}{N_1}{N_2} \sqsubseteq \ifelse{v_2}{N_3}{N_4}
  }
  \infrule[PC-Assert]{
    \forall\ST{\Gamma}.\refine{\Gamma}\land\pred\land\pred_1\Rightarrow\pred_2
    \andalso \Gamma;\pred\land\pred_1\p N_1 \sqsubseteq N_2
  }{
    \Gamma;\pred\p \assert{\pred_1};N_1 \sqsubseteq \assert{\pred_2};N_2
  }
  \caption{Precision relation of the cast terms (full version of Figure~\ref{fig:precision-term}).}
  \label{fig:precision-cast-term-}
\end{figure}

Figure~\ref{fig:precision-term-} and Figure~\ref{fig:precision-cast-term-} presents the full definition of the precision of terms $\seq{x}\p M_1\sqsubseteq M_2$ and the precision of cast terms $\Gamma;\pred\p N_1\sqsubseteq N_2$ respectively.

\begin{remark}
  The notation $\tilde{x}\setminus\{y\}$ denotes a sequence of variables that is created by removing $y$ from $\tilde{x}$.
\end{remark}

\begin{lemma}\label{lemma:type-subst}
  $\dom{\Gamma}\p\tau_1\sqsubseteq\tau_2$, $v_1 \sqsubseteq v_2$ and
  $\p v_1:\Gamma(x)$ imply
  $(\dom{\Gamma}\setminus\{x\})\p[v_1/x]\tau_1\sqsubseteq[v_2/x]\tau_2$.
\end{lemma}

\begin{proof}
  By induction on $\Gamma\p\tau_1\sqsubseteq\tau_2$.
  \begin{itemize}
    \item Case \rn{(Prec-Base)}.
    \infrule{
      \forall\dom{\Gamma},y.\pred_1\Rightarrow\pred_2
    }{
      \dom{\Gamma}\p \{y:B\mid\pred_1\} \sqsubseteq \{y:B\mid\pred_2\}
    }
    We can assume $y \neq x$ w.l.o.g by alpha renaming.

    If $\Gamma(x)$ is a function type, $x$ does not appear in $\pred_1$ or $\pred_2$. Therefore, the result follows trivially.

    If $\Gamma(x)$ is a base type, $v_1$ is a constant, and thus $v_1 = v_2$ follows from $v_1\sqsubseteq v_2$.
    Therefore, $\forall\dom{\Gamma}\setminus\{x\},y.[v_1/x]\pred_1\Rightarrow[v_2/x]\pred_2$ follows.

    \item Case \rn{(Prec-Fun)}.
    \infrule{
      \dom{\Gamma}\p \tau_3 \sqsubseteq \tau_5
      \andalso \dom{\Gamma},y\p\tau_4 \sqsubseteq \tau_6
    }{
      \dom{\Gamma}\p y{:}\tau_3\to\tau_4 \sqsubseteq y{:}\tau_5\to\tau_6
    }
    We can assume $y \neq x$ w.l.o.g by alpha renaming.
    From the induction hypothesis, the following holds.
    \begin{gather*}
      \dom{\Gamma}\setminus\{x\}\p[v_1/x]\tau_3\sqsubseteq[v_2/x]\tau_5 \\
      (\dom{\Gamma}\setminus\{x\}),y\p[v_1/x]\tau_3\sqsubseteq[v_2/x]\tau_5
    \end{gather*}
    Therefore, we obtain $\dom{\Gamma}\setminus\{x\}\p[v_1/x](y{:}\tau_3\to\tau_4)\sqsubseteq[v_2/x](y{:}\tau_5\to\tau_6)$ as expected.
  \end{itemize}
\end{proof}

\begin{lemma}\label{lemma:cast-term-subst}
  $\Gamma;\pred\p N\sqsubseteq N'$, $v \sqsubseteq v'$ and
  $\p v:\Gamma(x)$ imply
  $[v/x]\Gamma;[v/x]\pred\p [v/x]N \sqsubseteq [v'/x]N'$.
\end{lemma}

\begin{proof}
  By induction on $\Gamma;\pred\p N \sqsubseteq N'$.
  \begin{itemize}
    \item Case \rn{(PC-Const)} is trivial.
    \item Case \rn{(PC-Var)}.
    If $N = N' = x$, then $[v/x]N = v$ and $[v'/x]N' = v'$. Therefore, $[v/x]\Gamma;[v/x]\pred\p [v/x]N \sqsubseteq [v'/x]N'$ holds from $v\sqsubseteq v'$.
    Otherwise, $[v/x]\Gamma;[v/x]\pred\p [v/x]N \sqsubseteq [v'/x]N'$ trivially holds from \rn{(PC-Var)}.

    \item Case \rn{(PC-Lam)}.
    \infrule{
      \dom{\Gamma}\p \tau_1 \sqsubseteq \tau_2
      \andalso \Gamma,y:\tau_1;\pred\p N_1 \sqsubseteq N_2
    }{
      \Gamma;\pred\p\lambda y^{\tau_1}.N_1 \sqsubseteq \lambda y^{\tau_2}.N_2
    }
    We can assume $x \neq y$ w.l.o.g by alpha renaming.
    From the induction hypothesis, $([v/x]\Gamma),y:[v/x]\tau_1;[v/x]\pred\p [v/x]N_1\sqsubseteq[v'/x]N_2$ holds.
    Also, $\dom{\Gamma}\setminus\{x\}\p [v/x]\tau_1\sqsubseteq[v'/x]\tau_2$ holds from Lemma~\ref{lemma:type-subst}.
    Noting $\dom{[v/x]\Gamma}=\dom{\Gamma}\setminus\{x\}$,
    we obtain $[v/x]\Gamma;[v/x]\pred\p\lambda y^{[v/x]\tau_1}.[v/x]N_1 \sqsubseteq \lambda y^{[v'/x]\tau_2}.[v'/x]N_2$ as expected.

    \item Case \rn{(PC-Let)}.
    \infrule{
      \Gamma\p N_1\sqsubseteq N_3
      \andalso \Gamma,y:\tau_1;\pred\p N_2 \sqsubseteq N_4
    }{
      \Gamma;\pred\p \letin{y^{\tau_1}=N_1}N_2 \sqsubseteq \letin{y^{\tau_3}=N_3}N_4
    }
    We can assume $x \neq y$ w.l.o.g by alpha renaming.
    From the induction hypothesis, followings hold.
    \begin{gather*}
      [v/x]\Gamma;[v/x]\pred\p [v/x]N_1\sqsubseteq[v/x]N_3 \\
      ([v/x]\Gamma),y:[v/x]\tau_1;[v/x]\pred\p [v/x]N_2\sqsubseteq[v/x]N_4
    \end{gather*}
    Therefore,
    \begin{gather*}
      [v/x]\Gamma;[v/x]\pred\p(\letin{y^{[v/x]\tau_1}=[v/x]N_1}[v/x]N_2)
      \sqsubseteq (\letin{y^{[v/x]\tau_3}=[v/x]N_3}[v/x]N_4)
    \end{gather*}
    holds as expected.

    \item Case \rn{(PC-Fix)} and \rn{(PC-Let)} is similar to \rn{(PC-Lam)}.
    \item Case \rn{(PC-App)} and \rn{(PC-If)} is immediate from the induction hypothesis.

    \item Case \rn{(PC-Assert)}.
    \infrule{
      \forall\ST{\Gamma}.\refine{\Gamma}\land\pred\land\pred_1\Rightarrow\pred_2
      \andalso \Gamma;\pred\land\pred_1\p N_1 \sqsubseteq N_2
    }{
      \Gamma;\pred\p \assert{\pred_1};N_1 \sqsubseteq \assert{\pred_2};N_2
    }
    From the induction hypothesis, $[v/x]\Gamma;[v/x]\pred\land[v/x]\pred_1\p[v/x]N_1\sqsubseteq[v'/x]N_2$ holds.

    If $\Gamma(x)$ is a function type, then $\ST{[v/x]\Gamma}=\ST{\Gamma}$ and $\refine{[v/x]\Gamma} = \refine{\Gamma}$ holds. Also, $x$ does not appear in $\pred,\pred_1$ or $\pred_2$.
    Therefore, $\forall\ST{[v/x]\Gamma}.\refine{[v/x]\Gamma}\land[v/x]\pred\land[v/x]\pred_1\Rightarrow[v/x]\pred_2$ holds.

    If $\Gamma(x)$ is a base type, $v = v'$ holds from $v\sqsubseteq v'$ since $v$ must be a constant.
    Also, $\ST{[v/x]\Gamma}=\ST{\Gamma}\setminus\{x\}$ and $\refine{[v/x]\Gamma}=[v/x]\refine{\Gamma}$ holds.
    Therefore, $\forall\ST{[v/x]\Gamma}.\refine{[v/x]\Gamma}\land[v/x]\pred\land[v/x]\pred_1\rightarrow[v/x]\pred_2$ holds.

    Therefore, in either cases,
    \[
      \forall\ST{[v/x]\Gamma}.\refine{[v/x]\Gamma}\land[v/x]\pred\land[v/x]\pred_1\Rightarrow[v'/x]\pred_2
    \]
    follows as expected.
  \end{itemize}
\end{proof}

From now on, we abbreviate $\emptyset;\top\p N_1\sqsubseteq N_2$ as $N_1\sqsubseteq N_2$.

\vspace{10pt}
As we stated in the paper, we assume the following property for the primitive functions.
\begin{assumption}
  If $\evof{c,v_2}$ and $\evof{c,v_1}$ are both defined, then $v_1\sqsubseteq v_2$ implies $\evof{c,v_1}\sqsubseteq\evof{c,v_2}$.
\end{assumption}

\begin{lemma}\label{lemma:reduction-left}(Proposition~\ref{prop:reduction-left} in paper)
  Suppose $\p N_1:\tau$ and $\p N_2:\tau'$ for some $\tau$ and $\tau'$.
  Then, $N_1\sqsubseteq N_2$ and $N_1\reduce N_1'$ imply $N_2\reduce N_2'$ and $N_1'\sqsubseteq N_2'$.
\end{lemma}

\begin{proof}
  By induction on $N_1\reduce N_1'$.
  \begin{itemize}
    \item Case $\assert{\code{true}}; N_3 \reduce N_3$.
    From the inversion of $N_1 \sqsubseteq N_2$, it must be that $N_2 \equiv \assert{\code{true}}; N_4$ for some $N_4$ where $N_3\sqsubseteq N_4$. Therefore, $\assert{\code{true}};N_4 \reduce N_4$ holds.

    \item Case $(\lambda x^{\tau_1}.N_3)v_1 \reduce [v_1/x]N_3$.
    From the inversion of $N_1 \sqsubseteq N_2$, it must be that $N_2 \equiv (\lambda x^{\tau_2}.N_4)\,v_2$ for some $\tau_2$, $N_4$ and $v_2$ where $\tau_1\sqsubseteq\tau_2$, $v_1\sqsubseteq v_2$ and $x:\tau_1;\code{true}\p N_3\sqsubseteq N_4$.
    Also, from the inversion of $\p (\lambda x^{\tau_1}.N_3)\,v_1:\tau$,
    we obtain $\p v_1:\tau_1$.
    Therefore, by using Lemma~\ref{lemma:cast-term-subst}, $[v_1/x]N_3\sqsubseteq[v_2/x]N_4$ holds as expected.

    \item Case $(\fix{f^{\tau_1},x,N_1})v_1 \reduce [v_1/x,\fix{f^{\tau_1},x,N_1}/f]N_1$.
    From the inversion of $N_1 \sqsubseteq N_2$, it must be that $N_2 \equiv (\fix{f^{\tau_2},x,N_2})v_2$ where $\fix{f^{\tau_1},x,N_1} \sqsubseteq \fix{f^{\tau_2},x,N_2}$ and $v_1 \sqsubseteq v_2$.
    By using Lemma~\ref{lemma:cast-term-subst}, we obtain the following as expected.
    \[
      [v_1/x,\fix{f^{\tau_1},x,N_1}/f]N_1\sqsubseteq[v_2/x,\fix{f^{\tau_2},x,N_2}/f]N_2
    \]

    \item Case $c\,v_1\reduce\evof{c,v_1}$.
    From the inversion of $N_1 \sqsubseteq N_2$, it must be that
    $N_2 \equiv c\,v_2$ for some $v_2$ where $v_1\sqsubseteq v_2$ holds.
    Since $\p c\,v_2:\tau'$ holds, $\evof{c,v_2}$ is also defined.
    Using the property of $\evof{\cdot,\cdot}$, we obtain $\evof{c,v_1}\sqsubseteq\evof{c,v_2}$ as expected.

    \item Case $(\letin{x^{\tau_1}=v_1}N_3)\reduce[v_1/x]N_3$.
    From the inversion of $N_1\sqsubseteq N_2$, it must be that $N_2 \equiv \letin{x^{\tau_2}=v_2}N_4$ for some $\tau_2,v_2$ and $N_4$ where $v_1\sqsubseteq v_2$ and $x:\tau_1;\top\p N_3\sqsubseteq N_4$ holds.

    Also, we obtain $\p v_1:\tau_3$ and $x:\tau_3;\top\p N_3\sqsubseteq N_4$ for some $\tau_3$ from the inversion of $\p(\letin{x^{\tau_1}=v}N_3):\tau$.
    Using Lemma~\ref{lemma:cast-term-subst}, $[v_1/x]N_3\sqsubseteq[v_2/x]N_4$ holds as expected.

    \item Case $\ifelse{\top}{N_3}{N_4}\reduce N_3$.
    From the inversion of $N_1\sqsubseteq N_2$, it must be that
    $N_2\equiv\ifelse{\top}{N_5}{N_6}$ for some $N_5$ and $N_6$ where $N_3\sqsubseteq N_5$ and $N_4\sqsubseteq N_6$ holds.

    \item Case $\ifelse{\code{false}}{N_3}{N_4}\reduce N_3$ is similar to the previous case.

    \item Case $\letin{x^{\tau_1}=N_3}N_4\reduce\letin{x^{\tau_1}=N_3'}N_4$ where $N_3\reduce N_3'$.
    From the inversion of $N_1\sqsubseteq N_2$, it must be that $N_2\equiv(\letin{x^{\tau_2}=N_5}N_6)$ where $N_3\sqsubseteq N_5$ and $x:\tau_1;\pred\p N_4\sqsubseteq N_6$.
    From the induction hypothesis, $N_5\reduce N_5'$ and $N_3'\sqsubseteq N_5'$ holds for some $N_5'$. Therefore, $\letin{x^{\tau_2}=N_5}N_6\reduce\letin{x^{\tau_2}=N_5'}N_6$ and $\letin{x^{\tau_1}=N_3'}N_4\sqsubseteq\letin{x^{\tau_2}=N_5'}N_6$ holds as expected.

    \item Other cases are trivial from the induction hypothesis.
  \end{itemize}
\end{proof}

\begin{lemma}\label{lemma:reduction-right}(Proposition \ref{prop:reduction-right} in paper)
  Suppose $\p N_1:\tau$ and $\p N_2:\tau'$ for some $\tau$ and $\tau'$.
  Then, $N_1\sqsubseteq N_2$ and $N_2\reduce N_2'$ imply one of the following.
  \begin{itemize}
    \item $N_1\reduce N_1'$ and $N_1'\sqsubseteq N_2'$
    \item $N_1\reduce \blame$.
  \end{itemize}
\end{lemma}

\begin{proof}
  By induction on $N_2\reduce N_2'$.
  \begin{itemize}
    \item Case $\assert{\top};N_4\reduce N_4$.
    From the inversion of $N_1\sqsubseteq N_2$, it must be that $N_1\equiv(\assert{\pred};N_3)$ for some $\pred$ and $N_3$ where $\pred\Rightarrow\top$ and $N_3\sqsubseteq N_4$.
    Since $\pred$ is a closed predicate, $\pred\equiv\top$ or $\pred\equiv\code{false}$ holds.
    If $\pred\equiv\top$, then $\assert{\pred};N_3\reduce N_3$ and $N_3\sqsubseteq N_4$ holds as expected.
    Otherwise, $\assert{\pred};N_3\reduce\blame$ as expected.

    \item Case $(\lambda x^{\tau_2}.N_4)\,v_2 \reduce [v_2/x]N_4$.
    From the inversion of $N_1\sqsubseteq N_2$, it must be that $N_1\equiv(\lambda x^{\tau_1}.N_3\,v_1) \reduce [v_1/x]N_3$ for some $\tau_1,v_1$ and $N_3$ where $\tau_1\sqsubseteq\tau_2$, $v_1\sqsubseteq v_2$ and $x:\tau_1;\top\p N_3\sqsubseteq N_4$.
    By inverting $\p N_1:\tau$, we obtain $\p v_1:\tau_1$.
    Therefore, using Lemma~\ref{lemma:cast-term-subst}, $[v_1/x]N_3\sqsubseteq[v_2/x]N_4$ holds as expected.

    \item Case $(\fix{f^{\tau_2},x,N_4})\,v_2 \reduce [v_2/x,\fix{f^{\tau_2},x,N_4}/f]N_4$ is immediate using Lemma~\ref{lemma:cast-term-subst}.

    \item Case $c\,v_2\reduce\evof{c,v_2}$.
    From the inversion of $N_1 \sqsubseteq N_2$, it must be that
    $N_1 \equiv c\,v_1$ for some $v_1$ where $v_1\sqsubseteq v_2$ holds.
    Since $\p c\,v_1:\tau$ holds, $\evof{c,v_1}$ is also defined.
    Using the property of $\evof{\cdot,\cdot}$, we obtain $\evof{c,v_1}\sqsubseteq\evof{c,v_2}$ as expected.

    \item Case $\letin{x^{\tau_2}=v_2}N_4\reduce[v_2/x]N_4$.
    From the inversion of $N_1\sqsubseteq N_2$,
    it must be that $N_1\equiv\letin{x^{\tau_1}=v_1}N_3$ for some $\tau_1,v_1$ and $N_3$ where $\tau_1\sqsubseteq\tau_2$, $v_1\sqsubseteq v_2$ and $x:\tau_1;\top\p N_3\sqsubseteq N_4$.

    Also, $\p v_1:\tau_3$ and $x:\tau_3;\top\p N_3\sqsubseteq N_4$ holds for some $\tau_3$ from the inversion of $\p N_1:\tau$.
    Therefore, from Lemma~\ref{lemma:cast-term-subst}, we obtain $[v_1/x]N_3\sqsubseteq[v_2/x]N_4$ as expected.

    \item Other cases are trivial.
  \end{itemize}
\end{proof}

\begin{lemma}
  $\Gamma\sqsubseteq\Gamma'$, $\dom{\Gamma}\p M \sqsubseteq M'$ and $\Gamma;\pred\p M \leadsto N : \tau$ imply
  $\Gamma';\pred\p M' \leadsto N' : \tau'$, $\Gamma;\pred\p N \sqsubseteq N'$ and $\dom{\Gamma}\p\tau\sqsubseteq\tau'$ for some $N'$ and $\tau'$.
\end{lemma}

\begin{proof}
  By induction on the derivation of $\Gamma;\pred\p M\sqsubseteq M'$.
  \begin{itemize}
    \item Case \rn{(PM-Lam)}.
    \infrule{
      \dom{\Gamma}\p \tau_1\sqsubseteq\tau_1'
      \andalso \dom{\Gamma},x\p M_1 \sqsubseteq M_1'
    }{
      \dom{\Gamma}\p \lambda x{:}\tau_1.M_1 \sqsubseteq \lambda x{:}\tau_1'.M_1'
    }
    The derivation of $\Gamma;\pred\p\lambda x{:}\tau_1.M_1 \leadsto N:\tau$ must be of the following form.
    \infrule{
      \Gamma,x:\tau_1;\pred\p M_1 \leadsto N_1 :\tau_2
    }{
      \Gamma;\pred\p \lambda x{:}\tau_1.M_1 \leadsto \lambda x{:}\tau_1.N_1: x{:}\tau_1\to\tau_2
    }
    Let $\Gamma'$ such that $\Gamma\sqsubseteq\Gamma'$.
    Since $\Gamma,x:\tau_1\sqsubseteq\Gamma',x:\tau_1'$ holds,
    we can apply the induction hypothesis to obtain $\Gamma',x:\tau_1';\pred\p M_1' \leadsto N_1':\tau_2'$, $\Gamma,x:\tau_1;\pred\p N_1 \sqsubseteq N_1'$ and $\dom{\Gamma},x\p\tau_2\sqsubseteq\tau_2'$ for some $N_1'$ and $\tau_2'$.
    Therefore, $\Gamma';\pred\p \lambda x{:}\tau_1'.M_1' \leadsto \lambda x{:}\tau_1'.N_1':x{:}\tau_1'\to\tau_2'$ holds as expected.
    We conclude by noting $\dom{\Gamma}\p x{:}\tau_1\to\tau_2\sqsubseteq x{:}\tau_1'\to\tau_2'$.

    \item Case \rn{(PM-App)}.
    \infrule{
      \dom{\Gamma}\p M_1 \sqsubseteq M_1'
    }{
      \dom{\Gamma}\p M_1 \, x \sqsubseteq M_1' \, x
    }
    The derivation of $\Gamma;\pred\p M_1\,x\leadsto N:\tau$ must be of the following form.
    \infrule{
      \Gamma;\pred\p M_1 \leadsto N_1 :y{:}\tau_1\to\tau_2
      \andalso \Gamma(x)=\tau_3
      \andalso \Gamma;\pred\p \tau_3 \lesssim \tau_1 \leadsto N_2
    }{
      \Gamma;\pred\p M_1\,x \leadsto (\letin{y^{\tau_1}=N_2\,x}N_1\,y) :[x/y]\tau_2
    }
    Let $\Gamma'$ such that $\Gamma\sqsubseteq\Gamma'$.
    From the induction hypothesis, $\Gamma';\pred\p M_1' \leadsto N_1':y{:}\tau_1'\to\tau_2'$, $\Gamma;\pred\p N_1\sqsubseteq N_1'$ and $\dom{\Gamma}\p y{:}\tau_1\to\tau_2\sqsubseteq \tau'$ holds for some $N_1'$ and $\tau'$.
    By inverting $\dom{\Gamma}\p y{:}\tau_1\to\tau_2\sqsubseteq\tau'$, it must be that $\tau'\equiv y{:}\tau_1'\to\tau_2'$ for some $\tau_1'$ and $\tau_2'$ where $\dom{\Gamma}\p\tau_1\sqsubseteq\tau_1'$ and $\dom{\Gamma},y\p\tau_2\sqsubseteq\tau_2'$.
    Also, $\Gamma'(x)=\tau_3'$ and $\dom{\Gamma}\p\tau_3'\sqsubseteq\tau_3$ holds for some $\tau_3'$ from $\Gamma\sqsubseteq\Gamma'$.

    From Lemma~\ref{lemma:lesssim-relax} and Lemma~\ref{lemma:cast-precision}, we obtain $\Gamma';\pred\p\tau_3'\lesssim\tau_1'\leadsto N_2'$ for some $N_2'$ where $\Gamma;\pred\p N_2\sqsubseteq N_2'$ holds.
    Therefore,
    \[
      \Gamma';\pred\p M_1'\,x\leadsto(\letin{y^{\tau_1'}=N_2'\,x}N_1'\,y):[x/y]\tau_2'
    \]
    follows.
    We conclude by noting $\dom{\Gamma}\p[x/y]\tau_2\sqsubseteq[x/y]\tau_2'$ since the underlying types for $\tau_1$ and $\tau_3$ are the same, and the following.
    \[
      \Gamma;\pred\p (\letin{y^{\tau_1}=N_2\,x}N_1\,y)\sqsubseteq (\letin{y^{\tau_1'}=N_2'\,x}N_1'\,y)
    \]

    \item Case \rn{(PM-Let)}.
    \infrule{
      \dom{\Gamma}\p M_1 \sqsubseteq M_1'
      \andalso \dom{\Gamma},x\p M_2 \sqsubseteq M_2'
    }{
      \dom{\Gamma}\p (\letin{x^{\tau_1} = M_1} M_2) \sqsubseteq (\letin{x^{\tau_1'} = M_1'} M_2')
    }
    The derivation of $\Gamma;\pred\p \letin{x^{\tau_1}=M_1}M_2\leadsto N:\tau$ must be f the following form.
    \infrule{
      \Gamma;\pred\p M_1 \leadsto N_1 : \tau_1
      \andalso \Gamma,x:\tau_1\p M_2 \leadsto N_2 : \tau_2 \\
      \Gamma,x:\tau_1\p \tau_2\lesssim\tau \leadsto N_3
      \andalso \ST{\Gamma}\wf\tau
    }{
      \Gamma;\pred\p (\letin{x^{\tau_1}=M_1}M_2)
      \leadsto (\letin{x^{\tau_1}=N_1}\letin{y^{\tau_2}=N_2}N_3\,y) : \tau
    }
    Let $\Gamma'$ such that $\Gamma\sqsubseteq\Gamma'$.
    From the induction hypothesis, $\Gamma'\p M_1'\leadsto N_1':\tau_1'$ and $\Gamma',x:\tau_1';\pred\p M_2'\leadsto N_2':\tau_2'$ for some $N_1'$, $N_2'$, $\tau_1'$ and $\tau_2'$ where
    $\Gamma;\pred\p N_1\sqsubseteq N_1'$, $\Gamma,x:\tau_1;\pred\p N_2\sqsubseteq N_2'$, $\dom{\Gamma}\p\tau_1\sqsubseteq\tau_1'$ and $\dom{\Gamma},x\p\tau_2\sqsubseteq\tau_2'$.
    From Lemma~\ref{lemma:lesssim-relax-} and Lemma~\ref{lemma:cast-precision}, we obtain $\Gamma',x:\tau_-1';\pred\p\tau_2'\lesssim\tau\leadsto N_3'$ for some $N_3'$ where $\Gamma,x:\tau_1;\pred\p N_3\sqsubseteq N_3'$ holds, since $\Gamma,x:\tau_1\sqsubseteq\Gamma',x:\tau_1'$.
    Therefore,
    \begin{gather*}
      \Gamma;\pred\p (\letin{x^{\tau_1'}=M_1'}M_2')
      \leadsto(\letin{x^{\tau_1'}=N_1'}\letin{y^{\tau_2'}=N_2'}N_3'\,y):\tau
    \end{gather*}
    holds as expected. We conclude by noting
    \begin{gather*}
      \Gamma;\pred\p(\letin{x^{\tau_1}=N_1}\letin{y^{\tau_2}=N_2}N_3\,y)\sqsubseteq (\letin{x^{\tau_1'}=N_1'}\letin{y^{\tau_2'}=N_2'}N_3'\,y)
    \end{gather*}

    \item Case \rn{(PM-Fix)}.
    \infrule{
      \dom{\Gamma},f,x\p M_1 \sqsubseteq M_1'
      \andalso \dom{\Gamma}\p x{:}\tau_1\to\tau_2 \sqsubseteq x{:}\tau_1'\to\tau_2'
    }{
      \dom{\Gamma}\p \fix{f:(x{:}\tau_1\to\tau_2),x,M_1} \sqsubseteq \fix{f:(x{:}\tau_1'\to\tau_2'),x,M_1'}
    }
    The derivation of $\Gamma;\pred\p\fix{f:(x{:}\tau_1\to\tau_2),x,M_1}\leadsto N:\tau$ must be of the following form. (Note $\tau\equiv x{:}\tau_1\to\tau_2$)
    \infrule{
      \Gamma_1 = \Gamma,f:(x{:}\tau_1\to\tau_2), x:\tau_1
      \andalso \Gamma_1;\pred\p M_1 \leadsto N_1 :\tau_3
      \andalso \Gamma_1;\pred\p \tau_3 \lesssim \tau_2 \leadsto N_2
    }{
      \Gamma;\pred\p \fix{f:(x{:}\tau_1\to\tau_2),x,M_1}
      \leadsto \fix{f^{x{:}\tau_1\to\tau_2},x,\letin{y^{\tau_3}=N_1}N_2\,y}: x{:}\tau_1\to\tau_2
    }
    Let $\Gamma'$ such that $\Gamma\sqsubseteq\Gamma'$, and $\Gamma_1'=\Gamma,f:(x{:}\tau_1'\to\tau_2'),x:\tau_1'$.
    Then, $\Gamma_1\sqsubseteq\Gamma_1'$ holds.
    From the induction hypothesis, $\Gamma_1';\pred\p M_1'\leadsto N_1':\tau_3'$ holds for some $N_1'$ and $\tau_3'$ where $\Gamma_1;\pred\p N_1\sqsubseteq N_1'$ and $\dom{\Gamma},f,x\p\tau_3\sqsubseteq\tau_3'$.
    Also, $\dom{\Gamma},f,x\p\tau_2\sqsubseteq\tau_2'$ holds from the assumption of \rn{(PM-Fix)}.
    Using Lemma~\ref{lemma:lesssim-relax-} and Lemma~\ref{lemma:cast-precision}, $\Gamma_1';\pred\p\tau_3'\lesssim\tau_2'\leadsto N_2'$ holds for some $N_2'$ where $\Gamma;\pred\p N_2\sqsubseteq N_2'$ holds.
    Therefore,
    \begin{gather*}
      \Gamma';\pred\p\fix{f:(x{:}\tau_1'\to\tau_2'),x,M_1'}\leadsto
      \fix{f^{x{:}\tau_1'\to\tau_2'},x,\letin{y^{\tau_3'}=N_1'}N_2'\,y}:x{:}\tau_1'\to\tau_2'
    \end{gather*}
    holds as expected.
    We conclude by noting $\dom{\Gamma}\p x{:}\tau_1\to\tau_2\sqsubseteq x{:}\tau_1'\to\tau_2'$ and
    \begin{gather*}
      \Gamma;\pred\p \fix{f^{x{:}\tau_1\to\tau_2},x,\letin{y^{\tau_3}=N_1}N_2\,y}
      \sqsubseteq \fix{f^{x{:}\tau_1'\to\tau_2'},x,\letin{y^{\tau_3'}=N_1'}N_2'\,y}
    \end{gather*}

    \item Case \rn{(PM-If)}.
    \infrule{
      \dom{\Gamma}\p M_1 \sqsubseteq M_1'
      \andalso \dom{\Gamma}\p M_2 \sqsubseteq M_2'
    }{
      \dom{\Gamma}\p (\ifelse{x}{M_1}{M_2}) \sqsubseteq (\ifelse{x}{M_1'}{M_2'})
    }
    The derivation of $\Gamma;\pred\p\ifelse{x}{M_1}{M_2}:\tau$ must be of the following form.
    \infrule{
      \Gamma;\pred\p x : \{v:\bool\mid\pred'\} \\
      \Gamma;\pred\land x\p M_1 \leadsto N_1 : \tau_1
      \andalso \Gamma;\pred\land x\p \tau_1\lesssim\tau \leadsto N_3 \\
      \Gamma;\pred\land \lnot x\p M_2 \leadsto N_2 : \tau_2
      \andalso \Gamma;\pred\land \lnot x\p \tau_2\lesssim\tau \leadsto N_4
    }{
      \Gamma;\pred\p \ifelse{x}{M_1}{M_2} \\
      \leadsto \ifelse{x}{(\letin{y^{\tau_1}=N_1}N_3\,y)}{(\letin{y^{\tau_2}=N_2}N_4\,y)} : \tau
    }
    Let $\Gamma'$ such that $\Gamma\sqsubseteq\Gamma'$.
    We obtain the following from the induction hypothesis.
    \begin{gather*}
      \Gamma';\pred\land x\p M_1' \leadsto N_1' :\tau_1'
      \andalso \Gamma;\pred\land x\p N_1 \sqsubseteq N_1' \\
      \Gamma';\pred\land\lnot x\p M_2' \leadsto N_2' :\tau_2'
      \andalso \Gamma;\pred\land\lnot x\p N_2 \sqsubseteq N_2' \\
      \dom{\Gamma'}\p \tau_1\sqsubseteq\tau_1'
      \andalso \dom{\Gamma'}\p \tau_2\sqsubseteq\tau_2'
    \end{gather*}
    Using Lemma~\ref{lemma:lesssim-relax-}, $\Gamma';\pred\land x\p\tau_1'\lesssim\tau\leadsto N_3'$ and $\Gamma';\pred\land\lnot x\p\tau_2'\lesssim\tau\leadsto N_4'$ holds for some $N_3'$ and $N_4'$.
    From Lemma~\ref{lemma:cast-precision}, $\Gamma;\pred\land x\p N_3 \sqsubseteq N_3'$ and $\Gamma;\pred\land\lnot x\p N_4\sqsubseteq N_4'$ holds.
    Therefore, we obtain
    \begin{gather*}
      \Gamma';\pred\p\ifelse{x}{M_1'}{M_2'} \\
      \leadsto\ifelse{x}{(\letin{y^{\tau_1'}=N_1'}N_3'\,y)}{(\letin{y^{\tau_2'}=N_2'}N_4'\,y)}:\tau
    \end{gather*}
    as expected.
    We conclude by noting
    \begin{gather*}
      \Gamma;\pred\p
      \ifelse{x}{(\letin{y^{\tau_1}=N_1}N_3\,y)}{(\letin{y^{\tau_2}=N_2}N_4\,y)} \\
      \sqsubseteq
      \ifelse{x}{(\letin{y^{\tau_1'}=N_1'}N_3'\,y)}{(\letin{y^{\tau_2'}=N_2'}N_4'\,y)}
    \end{gather*}

    \item Case \rn{(PM-Annot)}.
    \infrule{
      \dom{\Gamma}\p M_1 \sqsubseteq M_1'
      \andalso \dom{\Gamma}\p \tau \sqsubseteq \tau'
    }{
      \dom{\Gamma}\p (M_1:\tau) \sqsubseteq (M_1':\tau')
    }
    The derivation of $\Gamma;\pred\p (M_1:\tau)\leadsto N:\tau$ must be of the following form.
    \infrule{
      \Gamma;\pred\p M_1 \leadsto N_1 : \tau_1
      \andalso \Gamma;\pred\p \tau_1 \lesssim \tau \leadsto N_2
    }{
      \Gamma;\pred\p (M_1:\tau) \leadsto (\letin{x^{\tau_1}=N_1}N_2\,x) : \tau
    }
    Let $\Gamma'$ such that $\Gamma\sqsubseteq\Gamma'$.
    From the induction hypothesis, $\Gamma';\pred\p M_1'\leadsto N_1':\tau_1'$ for some $N_1'$ and $\tau_1'$ where $\dom{\Gamma}\p\tau_1\sqsubseteq\tau_1'$ and $\Gamma;\pred\p N_1\sqsubseteq N_1'$.
    Using Lemma~\ref{lemma:lesssim-relax-} and Lemma~\ref{lemma:cast-precision}, we have $\Gamma';\pred\p\tau_1'\lesssim\tau'\leadsto N_2'$ for some $N_2'$ where $\Gamma;\pred\p N_2\sqsubseteq N_2'$ holds.
    Therefore, we obtain $\Gamma';\pred\p(M_1':\tau')\leadsto(\letin{x^{\tau_1'}=N_1'}N_2'\,x):\tau'$ as expected.
    We conclude by noting $\Gamma;\pred\p (\letin{x^{\tau_1}=N_1}N_2\,x) \sqsubseteq (\letin{x^{\tau_1'}=N_1'}N_2'\,x)$.
  \end{itemize}
\end{proof}

\end{document}

\end{document}